\documentclass[compsoc,conference,a4paper,10pt,times]{IEEEtran}

\usepackage[dvipsnames]{xcolor}

\usepackage[colorlinks=true,linkcolor=RubineRed,breaklinks=True,citecolor=RubineRed]{hyperref}
\usepackage{adjustbox}
\usepackage[linesnumbered,algo2e,ruled]{algorithm2e}
\usepackage{multirow}
\usepackage{booktabs,subcaption,dcolumn}
\usepackage{tikz}
\usepackage{enumitem}
\usepackage{tabularx}

\usepackage{pifont}
\usepackage{svg}
\usepackage{soul}
\usepackage[noadjust]{cite}

\usepackage[font=footnotesize]{caption}

\usepackage{amsmath, amsthm}
\usepackage{amsfonts, amssymb}
\usepackage{colortbl}
\newcommand{\cmark}{\ding{51}}%
\newcommand{\xmark}{\ding{55}}%

\usepackage{mathtools}
\newtagform{show_eq}{(Eq.\ }{)}  
\usetagform{show_eq}

\newcolumntype{?}{!{\vrule width 1.5pt}}

\newcommand{\takeaway}[1]{
    \vspace{1em}
    \noindent\fbox{%
        \parbox{\columnwidth}{%
            \textbf{Takeaway}: {#1}
        }%
    }
}

\newcommand*\halfcirc[1][0.7ex]{%
  \begin{tikzpicture}
  \draw[fill] (0,0)-- (90:#1) arc (90:270:#1) -- cycle ;
  \draw (0,0) circle (#1);
  \end{tikzpicture}}

\newcommand\revision[1]{%
  \bgroup
  \hskip0pt\color{blue!80!black}%
  #1%
  \egroup
}

\newcommand\revisionred[1]{%
  \bgroup
  \hskip0pt\color{red!80!black}%
  #1%
  \egroup
}

\newcommand\smallmath[1]{{\small$#1$}}
\newcommand\scriptmath[1]{{\scriptsize$#1$}}
\newcommand\dataset[1]{{\fontfamily{pcr}\selectfont {\scriptsize #1}}}

\theoremstyle{definition}
\newtheorem{definition}{Definition}
\newtheorem{requirement}{Req.}

\newcommand{\req}[1]{{\textsf{\footnotesize #1}}}
\newcommand{\overbar}[1]{\mkern 1.5mu\overline{\mkern-1.5mu#1\mkern-1.5mu}\mkern 1.5mu}

\def\BibTeX{{\rm B\kern-.05em{\sc i\kern-.025em b}\kern-.08em
    T\kern-.1667em\lower.7ex\hbox{E}\kern-.125emX}}
\begin{document}

\title{SoK: The Impact of Unlabelled Data in Cyberthreat Detection}

\author{\IEEEauthorblockN{Giovanni Apruzzese, Pavel Laskov, Aliya Tastemirova}
 \IEEEauthorblockA{\textit{Institute of Information Systems -- University of Liechtenstein} \\
 \{name.surname\}@uni.li
 }}

\maketitle

\pagestyle{plain}

\begin{abstract}
Machine learning (ML) has become an important paradigm for cyberthreat detection (CTD) in the recent years. A substantial research effort has been invested in the development of specialized algorithms for CTD tasks. 
From the operational perspective, however, the progress of ML-based CTD is hindered by the difficulty in obtaining the large sets of labelled data to train ML detectors. A potential solution to this problem are semisupervised learning (SsL) methods, which combine small labelled datasets with large amounts of unlabelled data. 

This paper is aimed at systematization of existing work on SsL for CTD and, in particular, on understanding the utility of unlabelled data in such systems. To this end, we analyze the cost of labelling in various CTD tasks and develop a formal cost model for SsL in this context. Building on this foundation, we formalize a set of requirements for evaluation of SsL methods, which elucidates the contribution of unlabelled data. We review the state-of-the-art and observe that no previous work meets such requirements. To address this problem, we propose a framework for assessing the benefits of unlabelled data in SsL. We showcase an application of this framework by performing the first benchmark evaluation that highlights the tradeoffs of 9 existing SsL methods on 9 public datasets. Our findings verify that, in some cases, unlabelled data provides a small, but statistically significant, performance gain.
This paper highlights that SsL in CTD has a lot of room for improvement, which should stimulate future research in this field.

\end{abstract}



\section{Introduction}
\label{sec:1-introduction}

Artificial intelligence and especially Machine Learning (ML) are one of the most important drivers of modern IT industry~\cite{Jordan:Machine}. Specifically, in cybersecurity they can play a crucial role in several tasks, such as vulnerability analysis~\cite{ghaffarian2017software}, security intelligence~\cite{nunes2016darknet}, and cyberthreat detection~(CTD)~\cite{Apruzzese:Deep}.

As pointed out by Sommer and Paxson \cite{Sommer:Outside}, the key to the tremendous success of the so-called \emph{supervised} ML methods is their ability to build models connecting the training input data with the known ground truth information, often referred to as ``labels''. It is well known, however, that obtaining the ground truth information in cybersecurity is a tough challenge~\cite{miller2016reviewer}. In computer vision, even a child can say whether a picture shows a cat or a dog. In natural language processing, a linguist -- albeit not always a layman -- can easily assign linguistic tags to (parts of) the text or assess the quality of its translation. Precise characterization of security events is orders of magnitude more difficult and brings supervised ML methods to a dilemma: they work best when a model is built using an extensive dataset, but the price of labelling such a dataset may be outrageously high. According to Miller et al.~\cite{miller2016reviewer}, an entire company can only afford to label 80 malware samples per day.

On the other hand, unlabelled data is abundant in cybersecurity. Petabytes of network traffic at many levels, monitoring and endpoints logs, as well as many other information sources (e.g., threat intelligence feeds~\cite{meier2018feedrank}) can be utilized for CTD tasks. Unlabelled data can be used in \emph{unsupervised} ML methods. 
For instance, clustering proved to be successful for some cybersecurity applications, e.g., analysis of malware families~\cite{bayer2009scalable}. However, such techniques can only address ancillary tasks, and cannot automate any detection (i.e., classification) mechanism without large amounts of labels that clearly distinguish benign from malicious samples (e.g.,~\cite{Chakraborty:EC2}).


\emph{Semisupervised} learning (SsL)~\cite{van2020survey} has a promise to deliver efficient ML classifiers that require \textit{small} amounts of labelled data by exploiting the information gained from \textit{large} sets of unlabelled data. For instance, in \textit{self learning} (e.g.,~\cite{koza2020two}), labelled data is mixed with unlabelled data to improve performance; similarly, in \textit{active learning}~\cite{gu2014active} an initial classifier trained on a small labelled dataset can be used to analyze a large set of raw data, and then `suggest' the most cost-effective samples to label.
The SsL setup resembles typical cybersecurity scenarios and, as a result, many works (e.g.,~\cite{noorbehbahani2020ransomware, sun2020effective, gabriel2016detecting}) proposed SsL solutions for diverse CTD tasks.


In a recent study from computer vision, Oliver et al.~\cite{oliver2018realistic} state, however, that ``\textit{the gap in performance between SsL and using only labelled data is smaller than reported}'', and that ``\textit{a classifier trained on a small labeled dataset with no unlabeled data can reach very good accuracy}''. These observations raise the question: is unlabelled data indeed beneficial to SsL and, if so, is it worth the effort? Unlabelled data is cheaper than labelling, but it is not completely `free'; e.g., its processing and storing costs are not negligible~\cite{huang2015empirical, ashadevi2009optimized}.

We researched previous works utilizing the combination of labelled and unlabelled data for CTD and found---surprisingly---that none of such works addressed the \textit{benefits} of unlabelled data in SsL. Identifying such benefits is not simple, because it requires the analysis of the key components that contribute to the development of SsL solutions. Therefore, to the best of our knowledge, it is still unclear whether SsL is advantageous for CTD. This is because prior works adopt evaluation protocols that are not standardized and do not allow to assess whether unlabelled data is cost-effective. Such immaturity serves as the main motivation for our SoK, and our specific goal is to promote deployment of SsL methods. 

To this end, we formalize a set of requirements derived by a systematic analysis of a realistic deployment of SsL methods in CTD. By following our requirements, it is possible to assess the impact of unlabelled data on the quality of SsL models. Moreover, we propose a novel evaluation framework that can be used to assess existing and future SsL methods in a research environment. Finally, we showcase an application of our framework to perform the first `benchmark' where we statistically verify the benefits of well-known SsL methods on 9 publicly available datasets for diverse CTD tasks.

Since the scope of this work is well beyond the traditional systematization of previous work, let us outline the \textbf{structure} and the \textbf{contributions} of this paper.

We begin with the presentation of the background and related research fields in §\ref{sec:background}. The main focus of this presentation is to illustrate the \textit{challenges of labelling data in CTD}, which is significantly more difficult than in other domains and, hence, motivates the search for solutions that can work with only a small amount of ground truth.. 

In §\ref{sec:SsL}, we formally present the general goal of SsL and analyze its cost structure. The main contributions of this section are an original cost model for SsL in CTD,
the definition of the benefit of unlabelled data in SsL, and \textit{the set of requirements} that must be met by research evaluations to ensure that such benefit can be claimed.

In §\ref{sec:sota}, we review the state-of-the-art on SsL for CTD, and analyze to what extent each paper meets the requirements identified in §\ref{sec:SsL}. The main conclusion of this analysis is that \emph{no prior work can satisfy all the requirements, and hence demonstrate the benefits of using unlabelled data}. This finding must not be interpreted as a deficiency of prior work; it merely highlights that no one ever questioned the utility of unlabelled data in the deployment of SsL for CTD.

As a first step towards such deployment, we present a new evaluation framework for SsL in §\ref{sec:framework} and demonstrate its application by assessing 9 well-known SsL methods using 9 datasets in §\ref{sec:demonstration}. All these contributions provide a constructive approach for assessing the utility of unlabelled data and promote the rollout of SsL for CTD.  

Finally, we discuss our findings (§\ref{sec:discussion}) and outline the conclusions of our study (§\ref{sec:conclusions}).
\section{Background and Related Work}
\label{sec:background}
Obtaining ground truth labels for training ML models is a well-known problem which motivates the investigation of SsL methods (e.g.,~\cite{owens2018audio, berthelot2019mixmatch}). However, with respect to other application domains of ML, the process of labelling in CTD is fundamentally harder, making SsL methods particularly attractive here. We elucidate all these labelling difficulties, representing the main motivation of our paper.

\subsection{Uniqueness of CTD with Respect to Labelling}
\label{ssec:ctd-vs-cv}
In some application domains of ML, there exist natural factors that facilitate acquiring labelled data for training ML models. Such factors can be \textit{data sharing} (e.g.,~\cite{chang2017revolt}), inherent \textit{low cost} of labelling (e.g., the popular CAPTCHAs~\cite{bursztein2011text}), or \emph{long-term usage} of labelled data (e.g., ImageNet was collected in 2009 and is still widely used today~\cite{you2018imagenet}).

All of these factors are not applicable to cybersecurity due to two intrinsic characteristics. First, the intrinsic confidentiality which strongly discourages data sharing. Second, the constant adaptation of attackers as well as the growth and the evolution of the environments being protected lead to the phenomenon known as \textit{concept drift}~\cite{jordaney2017transcend}, i.e., a fundamental change (gradual or abrupt) of the respective data generation processes.
This latter issue is crucial, as it conflicts with the underlying `iid' assumption\footnote{iid=independent and identically distributed random variables.} of ML and hence adversely affects its reliability in production environments~\cite{Sommer:Outside}. 

The peculiarities of CTD are especially clear in comparison to computer vision. In image recognition problems, the underlying ground truth is clear and stable. ``A cat will always be a cat, whereas a dog will always be a dog'' \cite{Pan:Transfer}, and a person can usually distinguish between two image classes very well~\cite{law2011human}, although specific applications may demand more informed opinions (e.g., physicians for cancer diagnoses~\cite{goldenberg2019new}). Moreover, after acquiring such labelled data, it is possible to apply \textit{data augmentation}~\cite{shorten2019survey} strategies to increase its effectiveness. For instance, adding some noisy pixels or mirroring the image allows one to create a new image that is different from the original but still has the same ground truth.

In contrast, all of the following can occur in CTD:
\begin{itemize}
    \item a malicious sample is benign elsewhere~\cite{Sommer:Outside};
    \item a malicious sample can be purposedly crafted to represent a benign sample~\cite{Apruzzese:Addressing};
    \item a benign sample today becomes a malicious sample tomorrow (the so-called `label shift'~\cite{lipton2018detecting}).
\end{itemize}
To aggravate the problem, verifying the ground truth of a sample is hard even for security experts, requiring further verifications~\cite{charlton2018measuring}. Finally, data augmentation is difficult to apply in CTD: for instance, changing a \textit{single} byte can turn many malicious samples into benign samples~\cite{Apruzzese:Evading}. 

Without loss of generality, we can state that a dataset that is usable for realistic CTD applications of ML must meet the following criteria~\cite{kaur2019systematic, cordero2021generating, noorbehbahani2015incremental}:
\begin{itemize}
    \item It must be \textit{large} enough to capture all the underlying characteristics of the environment to protect, and of the threats to defend against~\cite{Sommer:Outside}.
    \item The ratio of benign/malicious samples must be \textit{balanced} enough to allow efficient detection without generating excessive false alarms~\cite{yao2019msml}.
    \item It must have \textit{accurate ground truth}~\cite{charlton2018measuring}.
    \item It must be \textit{continuously updated}~\cite{jordaney2017transcend}.
\end{itemize}

These peculiarities make obtaining labelled datasets for CTD a tougher challenge than in other domains.


\subsection{Specific Labelling Issues in CTD}
\label{ssec:labelling}
We analyze the common procedures (and respective issues) for labelling data in CTD, split in three broad areas~\cite{Apruzzese:Deep, verma2019data}: Network Intrusion Detection (NID), Phishing Website Detection (PWD), Malware Detection (MD).

\textbf{Network Intrusion Detection.}
The detection of intrusions within a network perimeter can greatly benefit from ML~\cite{Buczak:Survey, Apruzzese:Deep} when labelled datasets are available. In the case of NID, such datasets contain samples providing network-related data. Well-known formats include full Packet Captures (PCAP) or Network Flows (NetFlows)~\cite{Bilge:Disclosure}. PCAP provides low-level information but it is not usable if the traffic is encrypted\footnote{Encrypted payloads also make labelling activities more difficult due to the impossibility of verifying the ground truth of a network packet.}; moreover, storing and analyzing PCAP is computationally expensive. In contrast, the NetFlow format mitigates these issues by providing a high-level overview of network communications between two endpoints while still enabling an appreciable detection performance~\cite{apruzzese2020deep, Bilge:Disclosure}. Other common data formats are DNS records for investigation of malicious domains~\cite{vekshin2020doh}, and SNMP for monitoring of specific hosts~\cite{fernandes2019comprehensive}. 
Regardless of the data-type, a common problem in NID is that every network is unique~\cite{Sommer:Outside, yao2019msml}. 
An anomalous behavior in one network may be normal in another network, hence preventing a reliable `transfer' of ML models. 
Moreover, obtaining accurate ground truth is tough for both \textit{benign} and \textit{malicious} samples. Most existing NID datasets used in research are created by infecting some machines in a controlled network environment with known malware, and capturing the traffic of the simulated network (e.g.,~\cite{CICIDS2018:Dataset, Garcia:CTU, UNSWNB15:Dataset}). 
Such approach is difficult to apply in reality. Verifying that a sample is truly legitimate requires ensuring that both hosts (the source and destination) `connected' by the specific traffic sample are not malicious. If one of these hosts is compromised via an unknown vulnerability then labelling all the traffic generated by such host as benign can lead to poisoning attacks~\cite{Apruzzese:Addressing}. 
Acquiring \textit{malicious} samples is also difficult, because it requires compromising \textit{real} machines with malware, hence exposing the network to external threats. Another problem is to consider all traffic originating from an infected machine is malicious, as some of its data may be generated by legitimate network activities (e.g., ARP messages).
All such challenges increase the difficulty of obtaining representative datasets for NID~\cite{mills2021practical}.

\textbf{Phishing Website Detection.}
Phishing attacks can be launched in various ways, e.g., via email or social networks~\cite{das2019sok}. In this paper, we focus on the detection of phishing websites because phishing usually involves luring the victim to enter some information on a (phishing) website. 
Attackers can easily create `squatting' websites that are difficult to detect~\cite{tian2018needle}, making them a rampant threat~\cite{Kettani:Threats}.
Datasets for PWD may include diverse data derived from, e.g., the URL (its length or the usage of some characters), the DNS record (a recent website is more likely to be a phishing hook), the HTML code (phishing websites have many pointers to external domains), and even the image of the landing page (most phishing hooks are similar to legitimate 
websites)~\cite{Corona:Deltaphish, Abdelhamid:Phishing, tan2016phishwho, mohammad2014predicting}, each having its pros and cons~\cite{marchal2018designing}.
The most common way to create labelled datasets for PWD is to use public lists of benign or malicious pages (e.g., AlexaTop or PhishTank). The webpage can be visited and the relevant information extracted to compose a given dataset~\cite{li2017phishbox}. In general, this process makes labelling for PWD easier than in NID, because verification is simple through expert knowledge~\cite{qabajeh2018recent}. Moreover, the ground truth of a website is independent on the target system (phishing page will `always' be malicious), enabling model transfer. However, such advantage presents some risks. For instance, the PWD embedded in the Google's Chrome web-browser was reverse engineered, allowing to craft phishing webpages that bypass detection~\cite{liang2016cracking}. Therefore, transferring ML models without any influx of new labelled data can result in unreliable PWD. 

\textbf{Malware Detection.}
The advent of data-driven solutions, such as ML, allowed the identification of malware variants which bypass traditional rule-based systems~\cite{Chakraborty:EC2}. Even commercial products leverage ML~\cite{cybersecurity2018machine}, which can be used for both static and dynamic MD~\cite{pham2018static, or2019dynamic}.
The complexity of labelling data for MD falls in between PWD and NID. 
Obtaining a large corpus of files (benign or malicious) for MD is not difficult per-se, as it can be done via public repositories. However, relying on such data without verifying the ground truth is risky. For instance, well-known marketplaces were recently found to contain malicious applications~\cite{kotzias2021did}. Treating such applications as \textit{benign} exposes to \textit{poisoning} attacks, and hence further verifications are required. Such verifications are, however, challenging: some web-services can automatically analyze an input (e.g., VirusTotal), but such services can disagree~\cite{verma2019data}, leading to unreliable results that require costly validation by experts~\cite{zhu2020measuring}. A possible workaround is using `collective wisdom' techniques that consider diverse antimalware engines~\cite{vsrndic2013detection}; recent works also propose to sanitize `noisy' labels in the training data~\cite{xu2021differential}.


\takeaway{Acquiring labelled data for CTD is challenging. To facilitate the development of ML-solutions, it is necessary to investigate approaches that can work when most of the data is not provided with the ground truth, such as SsL methods.}

\subsection{Focus of the Paper}
\label{ssec:focus}
\begin{figure}[!htbp]
    \centering
    \includegraphics[width=0.9\columnwidth]{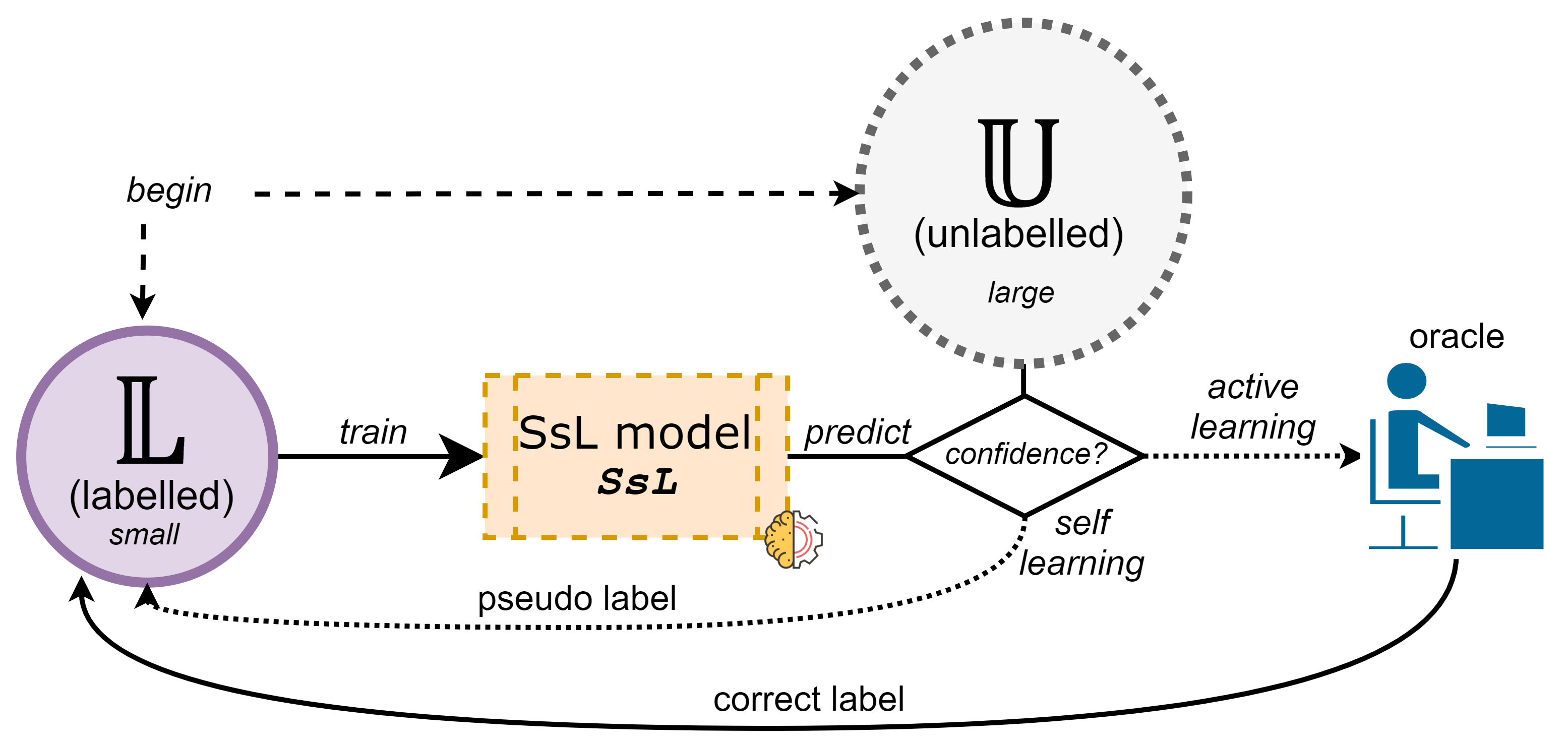}
    \caption{Active Learning and Pseudo Labelling.}
    \label{fig:methods}
\end{figure}

Combining labelled and unlabelled data to improve the proficiency of ML methods can be done in many ways. This SoK paper focuses on development of ML-systems for CTD, devised by using (i)~\textit{small} sets of labelled data (ii)~together with \textit{large} sets of unlabelelled data. Such settings align with the definition of `Semisupervised Learning' by Oliver et al.~\cite{oliver2018realistic}. To clarify our focus, we describe two exemplary techniques, illustrated in Fig.~\ref{fig:methods}: \textit{self learning via pseudo-labelling} (e.g.,~\cite{zhang2020semi}) and \textit{active learning via uncertainty-sampling} (e.g.,~\cite{rashidi2017android}), both associated with SsL~\cite{qiu2017flow, kumari2017semi, chandrasekaran2020exploring}.



In \textit{self learning}, the task of a ML model is to automatically learn from itself. In the specific case of \textit{pseudo labelling}, the intuition is to first train a ML model on a (small) set of labelled data ($\mathbb{L}$ in Fig.~\ref{fig:methods}), and then use such model to predict the label of a (large) set of unlabelled data ($\mathbb{U}$ in Fig.~\ref{fig:methods}). Such `pseudo-labels' are then used to retrain the model on a mixed dataset, containing both the correct labels of $\mathbb{L}$ and the pseudo-labels of $\mathbb{U}$. To avoid using potentially wrong labelled data, the retraining can be done by using only the pseudo-labelled samples with a high \textit{confidence} estimated by the model. 


In \textit{active learning}, the ML model interacts with an oracle (realistically, with a human) to improve its learning phase.
In the specific case of \textit{uncertainty sampling}, the idea is to first train a ML model on a (small) set of labelled data, $\mathbb{L}$, and then use such model to analyze a (large) set of unlabelled data, $\mathbb{U}$. The analysis is focused on `suggesting' to the oracle which samples in $\mathbb{U}$ should be correctly labelled to improve the performance, and the suggestion is based on the \textit{confidence} of the model on the samples in $\mathbb{U}$. Intuitively, the model can learn `more' from samples with a low confidence~\cite{zhang2021network}. The oracle then assigns the correct ground truth to such samples, which are inserted into $\mathbb{L}$ and used to retrain the model---using a correctly labelled dataset.

\subsection{Related Work}
\label{ssec:related}
Many learning paradigms focus on providing reliable models under the assumptions of scarce label availability; often, the term `Semisupervised Learning' is used to describe techniques that deviate from our definition. We summarize a few orthogonal areas to our work.

\textit{Federated learning}~\cite{rahman2020internet} aims to develop a `global' model by combining `local' models trained on small datasets. Despite some risks (e.g., poisoning~\cite{nguyen2020poisoning}) recent efforts applied it successfully even in privacy-sensitive settings~\cite{dayan2021federated}. However, federated learning makes no use of unlabelled data, and the final labelled dataset is huge.
    
\textit{Few-shot learning} has the goal of identifying unknown classes when only very few (or even zero~\cite{chen2017deep}) labels are available. For example, by finding the most relevant parameters of a baseline feature extractor, it is possible to generalize on unseen classes~\cite{ravi2016optimization}. These approaches---conceptually similar to `anomaly detection'---are thus tailored for detecting novel attacks, and showed promising achievements even in CTD (e.g.,~\cite{ale2020few}). However, they assume a large amount of initial labelled samples for the `known' classes. As an example, the authors of~\cite{marteau2021random} train their one-class classifiers on 80\% of the available samples for the `normal' class. Similarly, the ML-NIDS in~\cite{chen2017deep} can detect 14 `unseen' attacks, but is trained on over 200k samples spanning over 12 `known' classes.


\textit{Representation learning} focuses on finding the features that maximize the performance of a ML classifier~\cite{zhao2018xgbod}. It is possible to use unlabelled data to fine-tune such selection (e.g.,~\cite{fang2019semi, duarte2019semi, atzeni2018countering, Mirsky:Kitsune, mahindru2020feature}). Such procedures are ancillary to detection tasks, and hence outside our scope. As an example, the anomaly detector in~\cite{zhao2018xgbod} mixes a large set of unlabelled data with a small set of labels to identify the most representative features: however, once such features are identified, the experiments for the actual `detection' are performed by using 60\% of the (fully labelled) available data to train the classifier.

Finally, \textit{lifelong learning} aims to update ML models over-time~\cite{parisi2019continual, du2019lifelong}, which can be done by exploiting future (unlabelled) data streams; an assumption shared by active learning. 
However, sometimes the initial cost of labelling is neglected (this is the same problem as few-shot learning). An exemplary case is Tesseract~\cite{pendlebury2019tesseract}, focusing on time-aware MD. The idea of Tesseract is using, among others, active learning strategies to improve over time: despite increasing performance by 20\% with 700 additional labels, the initial deployment of Tesseract requires huge amounts of correct labels (i.e., more than 50K). Hence, some applications of active learning may have assumptions that deviate from ours.

We stress that our paper focuses on CTD. As such, any proposal that uses SsL for a different security-related task (e.g., fingerprinting~\cite{van2020flowprint}) is orthogonal to this paper.

%


\section{Semisupervised Learning for CTD}
\label{sec:SsL}
In contrast to obtaining verified labels, acquiring \textit{unlabelled} data for CTD is relatively straightforward.

Semisupervised Learning (SsL) aims to combine unlabelled with labelled data to devise ML models, which leads to the following question: ``what is the \textit{benefit} of unlabelled data?''. Only by answering this question it is possible to understand the role that SsL can play in CTD.

Let us begin our research by introducing SsL and its relationship with traditional supervised learning (SL).

\newcommand{\bbl}{\ensuremath{\overbar{\mathbb{L}}}}
\newcommand{\bl}{\ensuremath{\mathbb{L}}}
\newcommand{\bu}{\ensuremath{\mathbb{U}}}

\subsection{Introduction to Semisupervised Learning}
\label{ssec:intro_ssl}
Any ML model requires to \textit{learn} from data so that, after its deployment, it can efficiently analyze \textit{future} and unseen data, which we denote as $\mathbb{F}$. 
For those ML methods that require supervision (such as SL and SsL), the learning is done by means of \textit{labelled} data. To achieve the best performance, supervised models should be trained on a huge and fully labelled dataset, \bbl{}. Creating such \bbl{} may, however, be prohibitive.
In contrast, under the constraint of a limited labelling budget, $\mathcal{L}$, the amount of labelled data will be smaller, $\mathbb{L}$, and the resulting model may be inferior to the model that could have been built using \bbl{}.

SsL aims at bridging the gap between labelling effort and the model quality, by using $\mathcal{L}$ alongside a large \textit{unlabelled} dataset, $\mathbb{U}$, which is cheap to acquire. The resulting model should attain a better performance on $\mathbb{F}$ than a model built with same $\mathcal{L}$ but without using $\mathbb{U}$.

For instance, consider the two methods described in §\ref{ssec:focus}. In self-learning, the model should achieve a better performance after retraining on the `pseudo-labels' from $\mathbb{U}$ than the initial learner. In active learning, the model trained on the $\mathbb{L}$ with the `suggested' samples from $\mathbb{U}$ should outperform a model trained on a $\mathbb{L}$ derived from the same labelling budget $\mathcal{L}$, but without any `suggestions' derived by $\mathbb{U}$.

We now formally define the abovementioned scenarios.
Without loss of generality, any CTD task can be seen as 1+N \textit{classification} ML problem, where samples are either benign, or belong to one among N malicious classes. For simplicity, in the remainder we will consider a \textit{binary}-classification setting; all our considerations can be extended to cover also multi-classification settings.

Let $\mathcal{L}$ be a given labelling \textit{budget}. Let $\mathbb{L}$ be any \textit{labelled} dataset containing sample-label pairs, obtained by using $\mathcal{L}$. 
The composition of any labelled dataset is determined by its \textit{size}, \smallmath{|\!\cdot\!|}, and its \textit{class balance ratio}, $\rho(\cdot)$, which is a 1+N dimensional vector defining the distribution of its samples in percentages. Let $|\mathbb{L}|$ and $\rho(\mathbb{L})$ denote the size and class balance ratio\footnote{If N=1, an example is: $\rho(\mathbb{L}) \! = \! (60,\!40)$, i.e., 60\% of samples in $\mathbb{L}$ are benign, and 40\% are malicious.} of $\mathbb{L}$. 
Let $\overbar{\mathbb{L}}$, with $\rho(\mathbb{\overbar{L}})$, be a superset of $\mathbb{L}$.

Let $\mathbb{U}$ be an \textit{unlabelled} dataset containing samples 
of which the ground truth is not known, with $|\mathbb{L}| \! \ll \! |\mathbb{U}|$.

Finally, let $\mathbb{F}$ be another labelled dataset whose $|\mathbb{F}|$ and $\rho(\mathbb{F})$ should enable meaningful performance assessments. Because $\mathbb{F}$ represents future data, $\mathbb{F}$ $\cap$ $(\mathbb{\overbar{L} \cup \mathbb{U}}) \! = \! \varnothing$. Let \smallmath{\mu} be any performance \textit{metric}, e.g., accuracy or F1-score. 

\begin{figure}[!htbp]
    \centering
    \includegraphics[width=0.75\columnwidth]{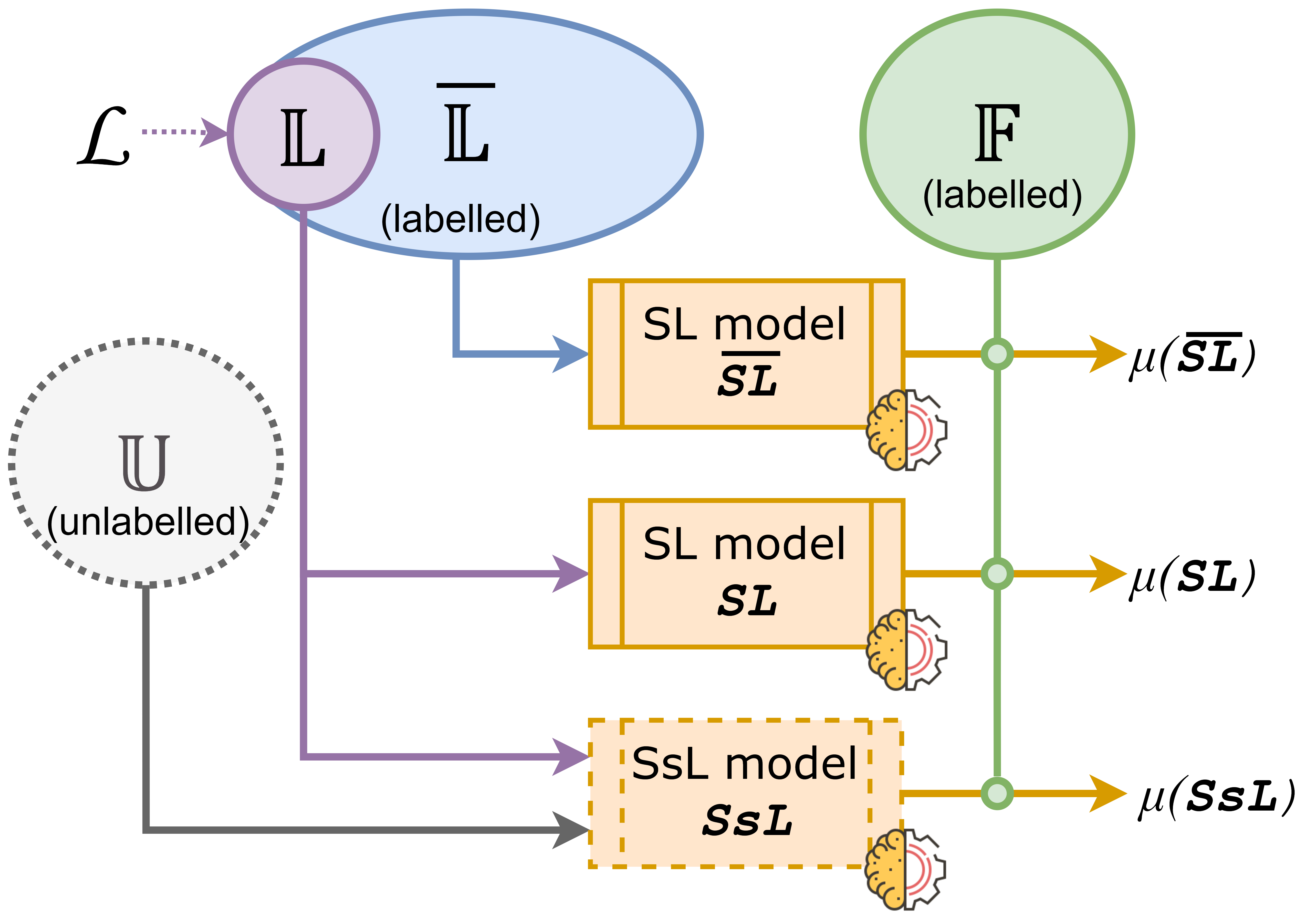}
    \caption{Semisupervised Learning w.r.t. Supervised Learning.}
    \label{fig:PUL}
\end{figure}

An illustration of the upcoming definitions is in Fig.~\ref{fig:PUL}.

A Supervised Learning (SL) method uses a labelled dataset to train a model that, after deployment, can predict the ground truth of unseen samples (i.e., in $\mathbb{F}$) obtaining a certain performance \smallmath{\mu(\cdot)}. 
To achieve optimal performance, the dataset used to train a model via SL should be comprehensive, that is, it should be adequately \textit{large} and \textit{balanced}. As a direct consequence, a model trained on a very small dataset will have a subpar performance. Let \smallmath{\overbar{SL}} be a model trained on $\overbar{\mathbb{L}}$, and \smallmath{SL} be a model trained on $\mathbb{L}$: if $|\mathbb{L}| \! \ll \! |\mathbb{\overbar{L}}|$ then \smallmath{\mu(SL)\!<\!\mu(\overbar{SL})}, irrespective of $\rho(\mathbb{L})$.
In the remainder, we assume\footnote{Finding the exact size and balance of a dataset that yield the best performance is a NP-hard problem.} that training \smallmath{\overbar{SL}} on $\mathbb{\overbar{L}}$ results in optimal \smallmath{\mu(\overbar{SL})}, and that $|\mathbb{L}| \! \ll \! |\mathbb{\overbar{L}}|$. 

We can now provide the following definition:

\begin{definition}
\label{def:SsL}
The \textit{goal} of a Semisupervised Learning (SsL) method is using $\mathbb{U}$ alongside any $\mathbb{L}$ obtained with $\mathcal{L}$ to devise a model \smallmath{SsL}.
After deployment, \smallmath{SsL} should predict the ground truth of the samples in $\mathbb{F}$ by achieving a performance \smallmath{\mu(SsL)} s.t.: \smallmath{\mu(SL)\!<\!\mu(SsL)\!\leq\!\mu(\overbar{SL})}.
\end{definition}

From Def.~\ref{def:SsL}, we derive that assessing the \textit{benefits} of SsL is linked with SL, because only by comparing\footnote{Doing this, however, requires that $\mathbb{L}$ is a strict subset of $\overbar{\mathbb{L}}$: otherwise (i.e., if $\mathbb{L} \! \cup \overbar{\mathbb{L}} \! \neq \overbar{\mathbb{L}}$) it would not be fair to compare {\scriptsize$\mu(SL)$} with {\scriptsize$\mu(\overbar{SL})$} and, consequently, with {\scriptsize$\mu(SsL)$}. This is because the performance difference may be due to the samples in $\mathbb{L}$ not included in $\overbar{\mathbb{L}}$.} \smallmath{SsL} with \smallmath{SL} (and \smallmath{\overbar{SL}}) the impact of $\mathbb{U}$ can be determined.

\subsection{Cost model of SsL for CTD}
\label{ssec:deployment}
\begin{figure}[!htbp]
    \centering
    \includegraphics[width=0.8\columnwidth]{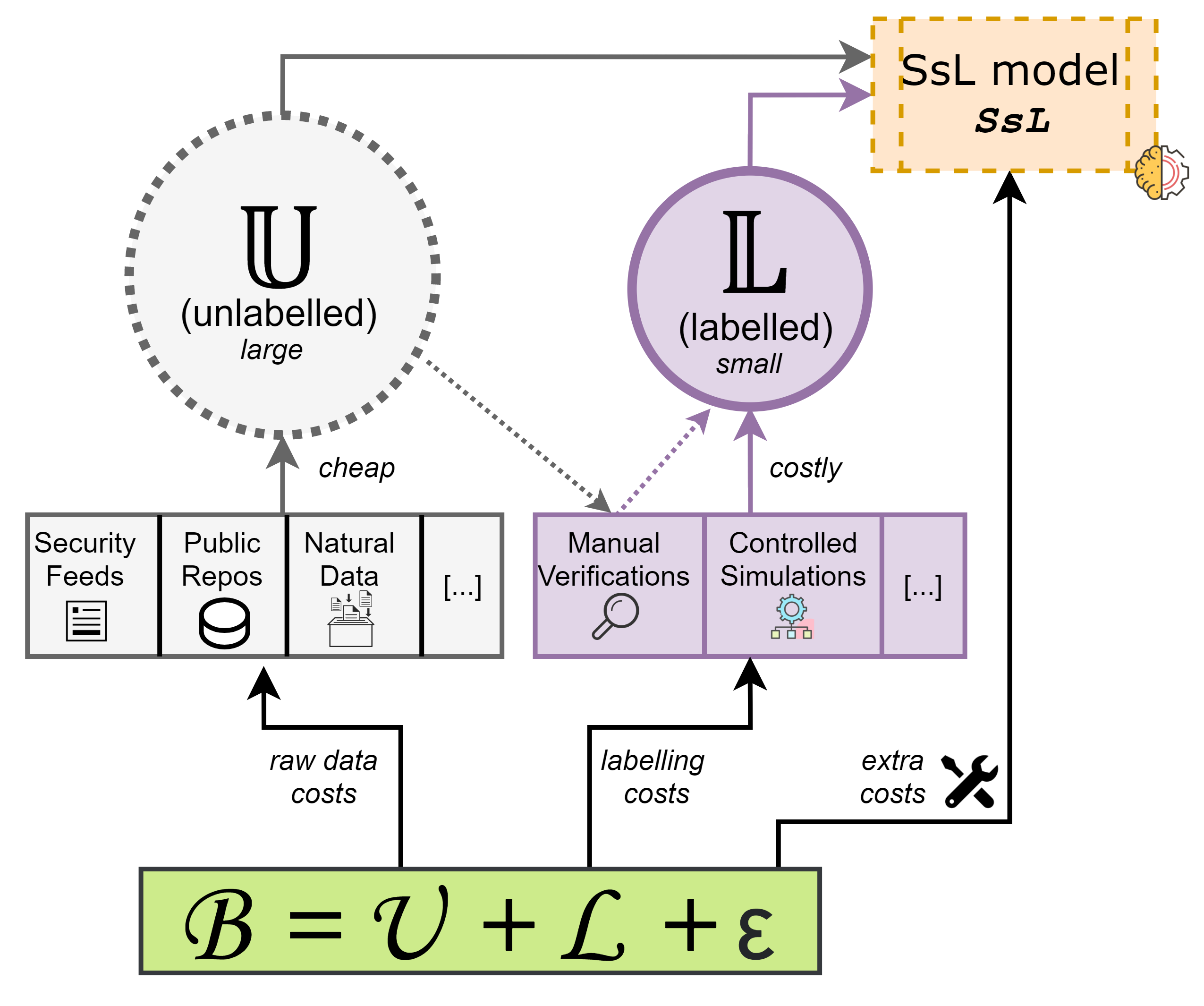}
    \caption{Propossed cost model for deployment of SsL methods.}
    \label{fig:cost-model}
\end{figure}

Let us interpret the abstract descriptions in §\ref{ssec:intro_ssl} from the perspective of a real organization willing to deploy a SsL solution for CTD. We provide an illustration in Fig.~\ref{fig:cost-model}.

In the real world, a resource \textit{investment} is required for integrating any new solution, and the decision to deploy such a solution depends on its estimated return on investment (ROI). Assessing the ROI demands the definition of a \textit{cost model} that allows practical comparisons to support operational decision making~\cite{chen2006evaluation}.

Huge amounts of data, $\mathbb{U}$, can be easily obtainable. An organization can have a rough idea about the nature of such data, but the impossibility of determining the ground truth without costly manual inspection makes it necessary to treat all its samples as unlabelled.

However, any SsL model necessitates at least a small amount of labelled data, $\mathbb{L}$. Such $\mathbb{L}$ can be acquired either before and independently from $\mathbb{U}$, or by using $\mathbb{U}$. For instance, $\mathbb{L}$ can be obtained by verifying uncertain labels, manually labelling existing data, or even by creating new labelled data via controlled simulations (§\ref{sec:background}).
In addition, like all ML-solutions, SsL models require ancillary operations (e.g., feature engineering, training and tuning~\cite{Apruzzese:Deep}).

We can hence define the \textit{cost model} for developing SsL solutions\footnote{TtBooK, we are the first to propose a SsL-specific cost model.}. Deployment of a semisupervised model \smallmath{SsL} requires to invest some \textit{budget}, $\mathcal{B}$.
Such budget can be seen as the sum of three elements $\mathcal{U}$, $\mathcal{L}$ and $\varepsilon$. Specifically:
\begin{itemize}
    \item $\mathcal{U}$ represents the investment for obtaining (and, if necessary, maintaining) the unlabelled data $\mathbb{U}$; 

    \item $\mathcal{L}$ is the investment for generating $\mathbb{L}$. Specifically, $\mathcal{L}$ is used to ensure that any sample $x \! \in \! \mathbb{L}$ has the \textit{correct} label.  Such $\mathcal{L}$ can be used all at once or at different times: for instance, it is possible to reserve a portion for subsequent labelling rounds that depend on $\mathbb{U}$ (e.g., active learning).
    
    \item \smallmath{\varepsilon(SsL)} is used for any \textit{extra} operation for developing \smallmath{SsL} that is not related to labelling the samples in $\mathbb{L}$. For instance, \smallmath{\varepsilon} can be used to (i) conduct preliminary analyses on $\mathbb{U}$, (ii) tune \smallmath{SsL}, (iii) process data, as well as for any (iv) computational costs.
\end{itemize}
By using $\mathcal{L}$, the organization eventually obtains a labelled dataset $\mathbb{L}$, whose composition (i.e., $\rho(\mathbb{L})$ and $|\mathbb{L}|$) depends on the cost of labelling each individual sample $x$ in $\mathbb{L}$; let $\mathcal{C}_x$ denote such \textit{cost}. 
We can express our cost model by formally defining $\mathcal{B}(SsL)$ with the following Equation:
\begin{equation}
    \mathcal{B}(SsL) = \mathcal{U} + \mathcal{L} + \varepsilon(SsL),~\text{where}~\mathcal{L} =  \sum_{x \in \mathbb{L}} \mathcal{C}_x
\label{eq:budget}
\end{equation}
Such budget $\mathcal{B}$ represents an \textit{investment} whose \textit{return} is based on the performance achieved by \smallmath{SsL} after deployment, that is, \smallmath{\mu(SsL)}. The ROI of any solution can be expressed as the ratio between its expected performance and its development budget~\cite{farris2010marketing}; in the case of SsL: ROI\smallmath{(SsL) \! = \! \mu(SsL) / \mathcal{B}(SsL)}.

We note, however, that \smallmath{\mu(SsL)} depends on `future' data which is, by definition, \textit{not} available to the organization.
It is hence necessary to conduct thorough experimental evaluations \textit{in advance} which can certify that: (i) a given \smallmath{SsL} (or its generative SsL method) will yield appreciable \smallmath{\mu(SsL)} when deployed in practice; and that (ii) allow comparisons of similar techniques by assessing their costs and benefits. 
If such evaluations are conducted on a dataset with a similar distribution of `future' data, then the resulting \smallmath{\mu(SsL)} will approximate the real deployment performance. In this case, it is possible to estimate the potential ROI of a SsL solution and facilitate informed decisions for their deployment. Nevertheless, after deployment, real organizations must regularly perform new evaluations to mitigate the likely concept drift (cf. §\ref{ssec:ctd-vs-cv}).

\takeaway{by investing $\mathcal{B}$, an organization can cheaply obtain $\mathbb{U}$ and compose a small $\mathbb{L}$ which are used to develop a model \smallmath{SsL}. Such model will only be developed (and deployed) if its estimated ROI is more beneficial than other solutions, i.e., if evaluations conducted \textit{in advance} show that \smallmath{SsL} outperforms solutions that require a lower or similar budget.}

\subsection{Requirements for Evaluation of SsL Methods}
\label{ssec:requirements}
Let us explain how to conduct evaluations that allow to estimate the ROI and, hence, gauge the benefits of SsL.

In research, evaluations of ML are done by means of fully labelled datasets, which should represent realistic scenarios and hence must be unbiased. Let $\mathbb{D}$ be one of such datasets\footnote{$\mathbb{D}$ can be subject to some preliminary preprocessing.}. From Def.~\ref{def:SsL}, a SsL model aims at \smallmath{\mu(SL)\!<\!\mu(SsL)\!\leq\!\mu(\overbar{SL})}. Therefore, the source dataset $\mathbb{D}$ must be used to to derive the four sets $\mathbb{F}$, $\mathbb{U}$, $\mathbb{L}$, and $\overbar{\mathbb{L}}$ necessary to compute such performance. To align the evaluation with real deployment use-cases, the creation of these sets must be done by taking $\mathcal{B}$ and, hence, $\mathcal{L}$ into account.
Note that $\mathbb{F}$ is only used to assess the performance on simulated `unknown' data, which is not available in advance. Therefore, using $\mathbb{F}$ to `cherry pick' the samples to put in $\mathbb{L}$ is prohibited~\cite{abu2012learning}.

We can now define the 7 requirements, applicable to any CTD task, that must be upheld to ensure unambiguous assessment of the \textit{benefits} of SsL methods. 

\begin{requirement}[\req{Lower Bound}]
\label{req:floor}
It is necessary to evaluate a lower bound model that only uses $\mathcal{L}$ and makes no use of $\mathbb{U}$. In other words, train a model \smallmath{SL} on $\mathbb{L}$ and evaluate its performance on $\mathbb{F}$ as \smallmath{\mu({SL})}.
To avoid bias\footnote{Labelling is not deterministic, and in realistic scenarios it is not possible to know \textit{in advance} the effectiveness of labelling.}, such $\mathbb{L}$ must be chosen by \textit{random} sampling from $\mathbb{D}$ subject to $\mathcal{L}$. 
\textbf{Motivation:} The \smallmath{SL} represents the true baseline\footnote{Such \scriptmath{SL} should aim at maximizing the gain from $\mathbb{L}$ (and hence $\mathcal{L}$), but it should not be overtuned (leading to overfitting and astronomical $\varepsilon$), nor it should be `misconfigured' to inflate results.}, allowing to assess the benefit (and, hence, compare the ROI) of any model \smallmath{SsL} that uses $\mathbb{L} \! + \! \mathbb{U}$. For instance, if \smallmath{\mu(SsL) \! \approx \mu(SL)} then there is no practical benefit in using the unlabelled data in $\mathbb{U}$; it may also be that \smallmath{\mu(SsL)\!<\!\mu(SL)}, meaning that using $\mathbb{U}$ is detrimental. 
\end{requirement}

\begin{requirement}[\req{Ablation Study}]
\label{req:random}
It is necessary to always consider a `vanilla' model \underline{\smallmath{SsL}} that uses $\mathbb{U}$ in a trivial way together with an $\mathbb{L}$ randomly sampled from $\mathbb{D}$. The aim is minimizing the degree of supervision\footnote{Depending on the context, there can be many ways of devising \scriptmath{\underline{SsL}}. As an example, in pseudo-labelling, a `trivial' way is using all pseudo-labels regardless of their confidence; whereas, in active learning, a `trivial' way is labelling randomly chosen samples (instead of those with least confidence).} involved by using $\mathbb{U}$. \textbf{Motivation:} The `vanilla' \underline{\smallmath{SsL}} allows to gauge (i) the smallest improvement provided by $\mathbb{U}$ via comparisons with \smallmath{SL}; and also (ii) the smallest \textit{cost} induced by using $\mathbb{U}$, because the randomness of $\mathbb{L}$ and the lack of supervision makes the corresponding $\varepsilon$ minimal. Moreover, \underline{\smallmath{SsL}} serves as a baseline for an \textit{ablation study}~\cite{berthelot2019mixmatch}, to simulate worst case scenarios in which any operation that relies on $\mathbb{U}$ to refinely compose $\mathbb{L}$ is not functional in practice.
\end{requirement}

\begin{requirement}[\req{Upper Bound}]
\label{req:ceiling}
It is necessary to train an upper bound model \smallmath{\overbar{SL}} on $\mathbb{\overbar{L}}$ and evaluate its performance on $\mathbb{F}$ as \smallmath{\mu(\overbar{SL})}. 
\textbf{Motivation:} The \smallmath{\overbar{SL}} serves to assess the performance achievable by augmenting $\mathbb{L}$ until it reaches \smallmath{\mathbb{\overbar{L}}}. Moreover, if \smallmath{\mu(SL)\!\approx\!\mu(\overbar{SL})}, then investing in $\mathcal{U}$ to develop any \smallmath{SsL} may not be worth it in the first place.
\end{requirement}

\begin{requirement}[\req{Statistical Significance}]
\label{req:stat}
It is necessary to verify the statistical significance of any evaluation result. \textbf{Motivation:} Evaluations of SsL involve a lot of randomness and uncertainties.
The huge search space for composing $\mathbb{L}$ from $\mathbb{D}$ may lead to erratic results: in some cases \smallmath{\mu(SsL)\!>\!\mu(SL)}, but the opposite can also be true. Moreover, sometimes \smallmath{\mu(SsL)\!\approx\!\mu(SL)} meaning that further tests are required to determine if \smallmath{SsL} is superior to \smallmath{SL}.
Hence, different draws of $\mathbb{L}$ (and, preferably, also of $\mathbb{\overbar{L}}$ and $\mathbb{F}$) must be assessed and conclusions must be drawn after statistically significant comparisons\footnote{Such efforts go beyond traditional `cross-validations'~\cite{pendlebury2019tesseract}.}. 
\end{requirement}

\begin{requirement}[\req{Transparency}]
\label{req:transparency}
It is necessary to ensure full transparency on the composition of $\mathbb{L}$, $\overbar{\mathbb{L}}$, $\mathbb{U}$ and $\mathbb{F}$. This implies: specifying \textit{size} of each dataset (both in absolute numbers, and with respect to $\mathbb{D}$); and the \textit{balance ratios} in terms of class composition. 
\textbf{Motivation:} Claiming that a given SsL method is effective when using only ``1\% of the data'' is not as enticing if $\mathbb{D}$ has 1M samples. Moreover, the balance ratio can significantly alter the results when small datasets are considered. This requirement also serves to estimate $\mathcal{L}$ and $\mathcal{U}$.
\end{requirement}

\begin{requirement}[\req{Reproducibility}]
\label{req:reproducibility}
Any evaluation must be supported with information that allow its reproducibility~\cite{baker2016reproducibility, pineau2020improving}. \textbf{Motivation:} aside from obvious reasons, it serves to approximate $\varepsilon$.
\end{requirement}

\begin{requirement}[\req{Multiple Settings}]
\label{req:datasets}
It is necessary to evaluate any model by considering multiple deployment settings, e.g., diverse datasets. \textbf{Motivation:} for practical deployments, the samples in $\mathbb{D}$ must resemble the true distribution at inference, and CTD scenarios can vary (cf. §\ref{sec:background}).
\end{requirement}

The intuition behind our requirements is that any model \smallmath{SsL} developed with a given SsL method must be compared against the three baselines 
of Regs.~\ref{req:floor}--\ref{req:ceiling}. By doing so, it is possible to measure the added value of \smallmath{SsL}, building on $\mathbb{U}$, potentially creating a better $\mathbb{L}$, but with a specific extra cost $\varepsilon$. It is implicit that any model-to-model comparison must be done under the assumptions of identical $\mathcal{L}$. 

We can now formally define the benefit of $\mathbb{U}$ in SsL.

\begin{definition}
\label{def:benefit}
Unlabelled data $\mathbb{U}$ used to develop any \smallmath{SsL} is \textit{beneficial} if it is shown that: (i)~\smallmath{\mu(SL)\!\ll\!\mu(\overbar{SL})}, and (ii)~ROI(\smallmath{SsL}) is better than both ROI(\smallmath{SL}) and ROI(\underline{\smallmath{SsL}}).
\end{definition}

Since \smallmath{SL} does not use $\mathcal{U}$, and \underline{\smallmath{SsL}} should minimize $\varepsilon$ (w.r.t. \smallmath{SsL}), it follows that \smallmath{\mu(SsL)} must be greater than \smallmath{\mu(SL)} and also greater or equal than \smallmath{\mu(\underline{SsL})}.
To justify deployment of a new SsL method, its evaluation must show that Def.~\ref{def:benefit} holds in different settings.

\takeaway{
Evaluating SsL in research requires to train and test (i) multiple models (ii) many times and in (iii) different scenarios, while reporting all details of the experimental setup. By meeting these requirements it is possible to assess the deployment benefits of a SsL method.}

\section{State-of-the-Art}
\label{sec:sota}
We analyze the state-of-the-art w.r.t. the proposed requirements (§\ref{ssec:requirements}), and provide a summary in Table~\ref{tab:sota}. Let us describe our methodology, and then discuss the main findings.
\newcommand{\al}{\cellcolor{gray!15}}
\newcommand{\ssl}{\cellcolor{white!40}}

\begin{table*}[!htbp]
    \centering
    \caption{State-of-the-Art of SsL for CTD w.r.t. our requirements. A `*' indicates a resource not available as of Sept 2021. Gray cells denote active learning. \textbf{All these works had a different scope than our paper, hence not meeting our requirements does not invalidate their contribution.}}

    \label{tab:sota}
    \resizebox{1.5\columnwidth}{!}{
    
        \begin{tabular}{c?c|c|c|c|c|c|c|c|c|c}
             
             \multirow{2}{*}{\textbf{Task}} & 
             \multirow{2}{*}{\textbf{Paper} \scriptsize{(1st Author)}} & 
             \multirow{2}{*}{\textbf{Year}} &
             \multirow{2}{*}{\begin{tabular}{c} \req{{\scriptsize Lower}} \\ \req{{\scriptsize Bound}} \end{tabular}} &
             \multirow{2}{*}{\begin{tabular}{c} \req{{\scriptsize Ablation}} \\ \req{{\scriptsize Study}} \end{tabular}} &
             \multirow{2}{*}{\begin{tabular}{c} \req{{\scriptsize Upper}}  \\ \req{{\scriptsize Bound}} \end{tabular}} &
             \multirow{2}{*}{\begin{tabular}{c} \req{{\scriptsize Stat.}} \\ \req{{\scriptsize Sign.}} \end{tabular}} &
             \multicolumn{2}{c|}{\req{{\scriptsize Transparency}}} &
             \multirow{2}{*}{\req{{\scriptsize Repr.}}} &
             \multirow{2}{*}{\req{Dataset}} \\
             \cline{8-9}
             & {} & {} & {} & {} & {} & {} & \textit{Labels} & \textit{Balance} & {} & {} 
             \\
             \toprule
             
                          \multirow{24}{*}{\rotatebox{90}{\textbf{Network Intrusion Detection}}} 
             & \al{} Li \cite{li2007active}  & 2007 &  \cmark  &  \cmark &  \xmark  &  \xmark  &  \cmark  &  \cmark  &  \halfcirc  &  \dataset{NSL-KDD} \\
            & \al{} Long \cite{long2008novel}  & 2008 &  \cmark  &  \cmark &  \xmark  &  \halfcirc  &  \cmark  &  \xmark  &  \halfcirc  &  \dataset{NSL-KDD} \\ 
            & \al{} G{\"o}rnitz \cite{gornitz2009active}  & 2009 &  \cmark  &  \cmark &  \xmark  &  \halfcirc  &  \cmark  &  \cmark  &  \xmark  &  Private \\
            & \al{} Seliya \cite{seliya2010active}  & 2010 &  \cmark  &  \cmark  &  \xmark  &  \xmark  &  \cmark  &  \cmark  &  \halfcirc  &  \dataset{NSL-KDD} \\
            & \ssl{} Symons \cite{symons2012nonparametric}  & 2012 &  \xmark  &  \cmark &  \cmark  &  \halfcirc  &  \cmark  &  \xmark  &  \xmark  &  \dataset{Kyoto2006} \\
            & \ssl{} Wagh \cite{wagh2014effective}  & 2014 &  \xmark  &  \xmark  &  \xmark  &  \xmark  &  \cmark  &  \cmark  &  \halfcirc  &  \dataset{NSL-KDD} \\
            & \al{} Noorbehbahani \cite{noorbehbahani2015incremental}  & 2015 &  \xmark  &  \halfcirc &  \cmark  &  \xmark  &  \cmark  &  \cmark  &  \halfcirc  &  \dataset{NSL-KDD}, Custom \\
            & \ssl{} Ashfaq \cite{ashfaq2017fuzziness}  & 2017 &  \xmark  &  \halfcirc  &  \xmark  &  \xmark  &  \cmark  &  \xmark  &  \halfcirc  &  \dataset{NSL-KDD} \\
            & \al{} Qiu \cite{qiu2017flow}  & 2017 &  \xmark  &  \halfcirc  &  \cmark  &  \xmark  &  \cmark  &  \cmark  &  \xmark  &  Custom \\
            & \al{} McElwee \cite{mcelwee2017active}  & 2017 &  \xmark  &  \halfcirc  &  \cmark  &  \xmark  &  \cmark  &  \xmark  &  \halfcirc  &  \dataset{NSL-KDD} \\
            & \al{} Kumari \cite{kumari2017semi}  & 2017 &  \cmark  &  \halfcirc &  \xmark  &  \xmark  &  \cmark  &  \xmark  &  \halfcirc  &  \dataset{NSL-KDD} \\
            & \al{} Yang \cite{yang2018active}  & 2018 &  \halfcirc  & \cmark &  \cmark  &  \xmark  &  \cmark  &  \xmark  &  \xmark  &  \dataset{NSL-KDD}, \dataset{AWID} \\
            
            & \al{} Gao \cite{gao2018novel}  & 2018 &  \cmark  &  \halfcirc  &  \xmark  &  \xmark  &  \cmark  &  \xmark  &  \xmark  & \dataset{NSL-KDD} \\
            
            & \ssl{} Shi \cite{shi2018self}  & 2018 &  \halfcirc  &  \halfcirc  &  \xmark  &  \xmark  &  \cmark  &  \xmark  &  \xmark  &  \dataset{NSL-KDD} \\
            & \ssl{} Yao \cite{yao2019msml}  & 2019 &  \halfcirc  &  \halfcirc  &  \cmark  &  \xmark  &  \cmark  &  \cmark  &  \halfcirc  &  \dataset{NSL-KDD} \\
            & \ssl{} Yuan \cite{yuan2019semi}  & 2019 &  \xmark  &  \halfcirc &  \xmark  &  \halfcirc  &  \cmark  &  \cmark  &  \halfcirc  &  \dataset{NSL-KDD} \\

            & \ssl{} Zhang \cite{zhang2020semi}  & 2020 &  \halfcirc  &  \xmark  &  \cmark  &  \halfcirc  &  \cmark  &  \xmark  &  \halfcirc  &  \dataset{NSL-KDD} \\
            
            & \ssl{} Hara \cite{hara2020intrusion} & 2020 &  \xmark  &  \halfcirc  &  \cmark  &  \xmark  &  \xmark  &  \xmark  &  \xmark  &  \dataset{NSL-KDD} \\
            
            & \ssl{} Ravi~\cite{ravi2020semisupervised} & 2020 &  \cmark  &  \xmark  &  \xmark  &  \xmark  &  \cmark  &  \xmark  &  \xmark  &  \dataset{NSL-KDD} \\

            & \ssl{} Gao \cite{gao2020saccos}  & 2020 &  \xmark  &  \cmark  &  \cmark  &  \cmark  &  \cmark  &  \cmark  &  \xmark  &  \dataset{NSL-KDD} \\

            & \ssl{} Li \cite{li2020enhancing}  & 2020 &  \xmark  &  \halfcirc  &  \cmark  &  \cmark  &  \cmark  &  \xmark  &  \halfcirc  &  \dataset{NSL-KDD}, Private \\
            & \ssl{} Zhang \cite{zhang2021network}  & 2021 &  \halfcirc  &  \halfcirc &  \xmark  &  \halfcirc  &  \xmark  &  \cmark  &  \halfcirc  &  \dataset{CICIDS2017}, \dataset{CTU13} \\
            
            & \ssl{} Liang \cite{liang2021fare}  & 2021 &  \cmark  &  \halfcirc &  \cmark  &  \halfcirc  &  \cmark  &  \cmark  &  \halfcirc  &  \dataset{NSL-KDD} \\

             \midrule
             
             \multirow{6}{*}{\rotatebox{90}{\parbox{40pt}{\centering\textbf{Phishing Detection}}}} 
             & \ssl{} Gyawali \cite{gyawali2011evaluating}  & 2011 &  \xmark  &  \cmark  &  \cmark  &  \xmark  &  \cmark  &  \cmark  &  \halfcirc  &  Private \\ 
            & \al{} Zhao \cite{zhao2013cost}  & 2013 &  \cmark  &  \cmark &  \cmark  &  \cmark  &  \xmark  &  \cmark  &  \cmark*  &  \dataset{DetMalURL} \\
            & \ssl{} Gabriel \cite{gabriel2016detecting}  & 2017 &  \halfcirc  &  \halfcirc &  \xmark  &  \xmark  &  \xmark  &  \xmark  &  \halfcirc  &  Private \\
            & \ssl{} Yang \cite{yang2017multi}  & 2017 &  \cmark  &  \halfcirc  &  \xmark  &  \xmark  &  \cmark  &  \cmark  &  \halfcirc  &  Private \\
            & \al{} Bhattacharjee \cite{bhattacharjee2017prioritized}  & 2017 &  \xmark &  \cmark &  \xmark  &  \halfcirc  &  \xmark  &  \xmark  &  \halfcirc  &  Private \\
            & \al{} Li \cite{li2017phishbox}  & 2017 &  \cmark  &  \cmark  &  \cmark  &  \halfcirc  &  \cmark  &  \cmark  &  \xmark  &  Custom \\

             \midrule

            \multirow{18}{*}{\rotatebox{90}{\parbox{40pt}\centering\textbf{Malware Detection}}}
             & \al{} Moskovitch \cite{moskovitch2008acquisition}  & 2008 &  \xmark  &  \cmark  &  \xmark  &  \halfcirc  &  \cmark  &  \cmark  &  \xmark  &  Custom \\
            & \ssl{} Santos \cite{santos2011semi}  & 2011 &  \xmark  &  \xmark  &  \cmark  &  \xmark  &  \cmark  &  \cmark  &  \halfcirc  &  Custom \\ 
            & \al{} Nissim \cite{nissim2012detecting}  & 2012 &  \xmark  &  \halfcirc &  \cmark  &  \halfcirc  &  \xmark  &  \xmark  &  \xmark  &  Private \\ 
            & \al{} Zhao \cite{zhao2012robotdroid}  & 2012 &  \xmark  &  \xmark &  \xmark  &  \xmark  &  \cmark  &  \cmark  &  \halfcirc  &  Private \\
            & \al{} Nissim \cite{nissim2014novel}  & 2014 &  \cmark  &  \cmark  &  \xmark  &  \halfcirc   &  \cmark  &  \cmark  &  \xmark  &  Custom \\
            & \al{} Zhang\cite{zhang2015update}  & 2015 &  \halfcirc  &  \halfcirc  &  \xmark  &  \xmark  &  \cmark  &  \cmark  &  \xmark  &  Private \\
            & \al{} Nissim \cite{nissim2016aldroid}  & 2016 &  \xmark  &  \cmark &  \cmark  &  \halfcirc  &  \cmark  &  \cmark  &  \halfcirc  &  Custom \\
            & \al{} Ni \cite{ni2016findmal}  & 2016 &  \cmark  &  \cmark  &  \xmark  &  \halfcirc  &  \cmark  &  \cmark  &  \halfcirc  &  Private \\
            & \ssl{} Chen \cite{chen2017poster}  & 2017 &  \cmark  &  \cmark &  \xmark  &  \halfcirc  &  \xmark  &  \xmark  &  \halfcirc  &  Private \\
            & \al{} Rashidi \cite{rashidi2017android}  & 2017 &  \xmark  &  \cmark &  \cmark  &  \halfcirc  &  \cmark  &  \cmark  &  \xmark  &  \dataset{Drebin} \\
            & \ssl{} Fu \cite{fu2019malware}  & 2019 &  \cmark  &  \cmark  &  \xmark  &  \xmark  &  \cmark  &  \xmark  &  \halfcirc  &  Private \\
            & \al{} Irofti \cite{irofti2019malware}  & 2019 &  \halfcirc  &  \halfcirc  &  \xmark  &  \halfcirc  &  \xmark  &  \xmark  &  \cmark  &  \dataset{DREBIN,EMBER} \\ 
            & \al{} Pendlebury \cite{pendlebury2019tesseract} & 2019 & \xmark & \xmark & \cmark & \halfcirc & \cmark & \cmark & \cmark & \dataset{AndroZoo} \\
            & \ssl{} Sharmeen \cite{sharmeen2020adaptive}  & 2020 &  \cmark  &  \halfcirc  &  \xmark  &  \halfcirc  &  \cmark  &  \cmark  &  \halfcirc  &  \dataset{Drebin}, \dataset{AndroZoo} \\
            & \al{} Chen \cite{chen2020malware}  & 2020 &  \halfcirc  &  \halfcirc  &  \cmark  &  \xmark  &  \cmark  &  \cmark  &  \halfcirc  &  \dataset{MCC} \\
            & \ssl{} Koza \cite{koza2020two}  & 2020 &  \cmark  &  \halfcirc  &  \cmark  &  \halfcirc  &  \cmark  &  \xmark  &  \cmark  &  Private \\
            
            & \ssl{} Noorbehbahani \cite{noorbehbahani2020ransomware}  & 2020 &  \cmark  &  \xmark &  \xmark  &  \halfcirc  &  \cmark  &  \cmark  &  \xmark  &  \dataset{AndMal17} \\ 
            & \ssl{} Li \cite{li2021semi}  & 2021 &  \xmark  &  \halfcirc  &  \xmark  &  \halfcirc  &  \cmark  &  \xmark  &  \halfcirc  &  \dataset{FalDroid,DREBIN,Genome} \\
            & \ssl{} Liang \cite{liang2021fare}  & 2021 &  \cmark  &  \halfcirc &  \cmark  &  \halfcirc  &  \cmark  &  \cmark  &  \halfcirc  &  Custom \\

            

             \bottomrule
        \end{tabular}

    }
\end{table*}

\subsection{Methodology}
\label{ssec:methodology}
To assess the extent to which existing works meet our requirements, we perform a systematic literature survey. Such process is organized in three phases: \textit{search}, \textit{screening}, \textit{investigation}. To reduce bias, all phases involved two researchers who worked \textit{independently}, and who discussed their individual findings in weekly meetings.

\subsubsection{Search} 
We first searched for all literature linking (even remotely) SsL with CTD  in well-known scientific repositories. Such repositories include IEEE Xplore, Google Scholar, and ACM Digital Library; but we also considered the proceedings of top security conferences.
In particular, we searched for the following keywords:

{\footnotesize
\begin{gather*}
    (semi\textnormal{-}supervised \lor semisupervised \lor semi~supervised \lor active) \\
    \land \\
    (network \lor malware \lor phishing \lor intrusion)
\end{gather*}
} which had to be included either in the title or in the abstract.
Any work that was not peer-reviewed was excluded, and we looked for papers published after 2007.
The results of such search formed an initial corpus of papers, which was further extended with all papers that either cited, or were cited by, a given work (and that included same or similar keywords); as well as with papers that the authors autonomously found during their daily duties (e.g., reviewing).

\subsubsection{Screening}
After obtaining the corpus of candidate papers, we studied them with the intent of determining which papers fall within our scope. By referring to §\ref{ssec:focus}, a paper had to meet three criteria: (i) focus on CTD, (ii) using unlabeled data, (iii) in combination with \textit{small} sets of labelled data\footnote{The term `Semisupervised Learning' has many meanings (§\ref{ssec:related}).}. After several discussions between the two researchers, this phase resulted in the set of 48 papers reported in Table~\ref{tab:sota}. To the best of our knowledge, Table~\ref{tab:sota} represents the current state-of-the-art of SsL for CTD. 


\subsubsection{Investigation}
Those papers that met all the inclusion criteria were then further analyzed, with the goal of assessing their compliance with the proposed set of requirements. The results are in Table~\ref{tab:sota}, which is organized as follows.
We distinguish the papers on the basis of the three main CTD areas of interest (NID, PWD, MD); cells with a gray background denote papers that specifically consider `active learning' approaches. For each paper we compare it with our requirements: a \cmark~(resp.~\xmark) denotes that a requirement is met (or not).
\begin{itemize}
    \item For Req.~\ref{req:floor}, we use \halfcirc\ if it is not explicitly mentioned that $\mathbb{L}$ was randomly drawn, and \xmark\ when either the \smallmath{SL} is missing, or when such \smallmath{SL} uses a different $\mathcal{L}$.
    \item For Req.~\ref{req:random}, we use \cmark\ if the paper considers a SsL model that is completely unbiased and can serve as an ablation study; \halfcirc\ if the SsL models are trained on a unbiased random $\mathbb{L}$, but cannot serve as an ablation study due to `overuse' of $\mathbb{U}$; and \xmark\ if the provided information is insufficient to determine the absence of bias in $\mathbb{L}$.
    \item Req.~\ref{req:ceiling} is binary, but we also consider as \cmark\ when \smallmath{SsL} is trained on a very large set of correct labels.
    \item Req.~\ref{req:stat} we use \halfcirc\ if only some `cross-validation' is performed, \cmark\ if statistical comparisons are made or mentioned, and \xmark\ otherwise.
    \item For Req.~\ref{req:transparency}, we report two columns: `\textit{Labels}' denotes whether the provided information allows to determine the actual number of labelled samples used to train and test all the considered models; `\textit{Balance}' denotes whether the balancing ratios are clearly specified.
    \item For Req.~\ref{req:reproducibility}, \xmark\ denotes if the provided information is insufficient for reproduction; \cmark\ if the source code is open; and \halfcirc\ if only intermediate information is provided.
\end{itemize}
In the last column we report the datasets used in each paper: here, `Private' means that the data was never made available, whereas `Custom' means that it was composed in-house via public sources, but that the actual samples cannot be recovered (i.e., it is not possible to retrieve the public feeds of past years).

\subsection{Findings}
\label{ssec:findings}
By observing Table~\ref{tab:sota}, we derive the following.
\begin{enumerate}
    \item No one fits all: no paper meets all requirements. 
    \item Few compare their SsL methods with a lower bound.
    \item Worst case scenarios are rarely covered.
    \item Lack of statistically significant comparisons, preventing any certification of the final results.
    \item Poor reproducibility and limited datasets, which is a known trend in ML research~\cite{hutson2018artificial}.
\end{enumerate}
Although no paper meets all our requirements, there are some good efforts. 
Remarkably, Zhao et al.~\cite{zhao2013cost} meet almost all requirements (with the exception of a single dataset, and their implementation not being available today), and are the only ones to mention a statistical comparison via a student t-test; however, we were not able to infer how the starting dataset was split in training and testing. We also praise the work in~\cite{koza2020two}, but it lacks a vanilla \smallmath{\underline{SsL}} for ablation studies (due to fine tuning of confidence thresholds), and only a limited cross-validation is performed. 
Noteworthy is also~\cite{liang2021fare}, whose authors fairly evaluate different SsL techniques (although none of these can be considered as an ablation study) for two CTD tasks (NID and MD), but each on a single dataset.

The situation portrayed by Table~\ref{tab:sota} does not imply that all past works are wrong or flawed: these papers are published in high quality venues, and we acknowledge their significance. 
On the contrary, the true message of Table~\ref{tab:sota} is highlighting the \textit{immaturity} of the state-of-the-art with respect to realistic deployments of SsL. No attention has been given to systematic assessments of the benefits provided by unlabelled data in SsL. 

\textbf{The case of active learning.}
We observe that many papers in Table~\ref{tab:sota} use active learning methods, most of which in lifelong learning settings. In these cases, considering a model trained via random draws from $\mathbb{U}$ (instead of `active' suggestions) can simultaneously meet both Req.~\ref{req:floor} and~\ref{req:random}. This is notably done by Gornitz et al.~\cite{gornitz2009active} (despite not meeting Req.~\ref{req:ceiling}). In contrast, Pendlebury et al.~\cite{pendlebury2019tesseract} apply active learning by labelling \textit{all} samples with confidence below 1\%, which results in 700 samples, and the improvement is shown against a model that does not make use of any additional label, leading to an unfair comparison.
The evaluation protocol of these and similar papers is, however, \textit{legitimate}. They operate under the assumption that the ML model is already trained and deployed, meaning that unlabelled data will naturally occur. In such conditions, the focus is not on ``determining the benefits of unlabelled data for ML deployment'' (which is our focus), but rather on ``how to maximize the performance of an existing ML system with additional unlabelled data streams''. Both~\cite{pendlebury2019tesseract} and~\cite{gornitz2009active} achieve that. Nonetheless, to the best of our knowledge (semi)\textit{supervised} ML systems are not widely deployed (yet) in CTD, demanding further investigations of their potential benefits \textit{in advance}.

\textbf{Relationship with other domains.}
There are several studies that expose evaluation issues of ML methods, and Dehgani et al.~\cite{dehghani2021benchmark} invite devising specific guidelines. However, within the context of SsL, existing proposals are \textit{not applicable} to CTD. For instance, Oliver et al.~\cite{oliver2018realistic} suggest transferring models between different datasets: this may not be feasible for CTD because datasets contain divergent feature sets and model transferring can be a recommendation at best, and not a requirement. Similarly,~\cite{musgrave2020metric} mention random sampling, but do not emphasize the statistical significance which is crucial for SsL in CTD due to the huge search space to extract a small $\mathbb{L}$ from a huge $\mathbb{D}$. This is less of a problem in, e.g., Computer Vision, where most datasets (e.g., \dataset{CIFAR}) have been used thousands of times and benchmarks results are well-known; moreover, the corresponding community is more open to source code disclosure. Because of these reasons, it is crucial to establish a specific set of requirements for CTD applications of SsL.

Regardless, some of our requirements are not met also by relevant works. An exemplary case, which exploits data augmentation in \dataset{CIFAR}, is MixMatch~\cite{berthelot2019mixmatch}. Here, no lower bound \smallmath{SL} is considered: as a matter of fact, they only report the results of the upper bound \smallmath{\overbar{SL}} trained on 50K samples (i.e., most of \dataset{CIFAR}). We do acknowledge that \dataset{CIFAR} is well-known and performance on small subsets can be easily assessed, but a fair comparison requires to evaluate such performance by using the same settings (i.e., same $\mathbb{L}$ and same classifier).

\takeaway{the state-of-the-art does not allow to assess the benefits of SsL methods. Such immaturity is due to the lack of a rigorous evaluation protocol for SsL in CTD.}



\section{Proposed Evaluation Framework}
\label{sec:framework}
As a constructive step forward, we present CEF-SsL, an original Cybersecurity Evaluation Framework for SsL, which meets all the requirements in §\ref{ssec:requirements}.

CEF-SsL aims to provide a practical assessment of SsL methods by using any fully labelled dataset, while simultaneously considering the deployment budget $\mathcal{B}$ and ensuring the statistical significance of the results. Because $\mathbb{U}$ can be easily obtained, CEF-SsL assumes that $\mathcal{U}$ is fixed and, hence, plays no role in practical comparisons.
CEF-SsL has four \textit{inputs}: 
\begin{itemize}
    \item $\mathbb{D}$ represents a large and fully labelled dataset. Such $\mathbb{D}$ can be either: (i) openly accessible; or (ii) created ad-hoc via simulations, well-known security feeds, or by manually labelling real data. CEF-SsL assumes that the labels in $\mathbb{D}$ are verified, i.e., all samples have the correct ground truth, which is the typical assumption in ML research. CEF-SsL uses $\mathbb{D}$ as basis\footnote{$\mathbb{D}$ is assumed to be already preprocessed, and it can be a subset of an existing dataset, but the selection must be unbiased.} to compose the four datasets required for a practical evaluation $\mathbb{L}$, $\mathbb{\overbar{L}}$, $\mathbb{U}$, and $\mathbb{F}$.
    
    \item $\mathcal{L}$ is the labelling budget which is used to to compose $\mathbb{L}$ (cf Eq.~\ref{eq:budget}). 
    CEF-SsL assumes that $\mathcal{L}$ is fixed for the entire simulation. Variations of $\mathcal{L}$ imply different scenarios (hence, different $\mathbb{L}$) which lead to unfair comparisons among models. If necessary, CEF-SsL can be applied again on different values of $\mathcal{L}$.
    
    \item {\footnotesize$\overrightarrow{\text{ML}}$} denotes an \textit{array of ML methods}. Such array must contain the specifics to devise all the baseline models (\smallmath{SL}, \smallmath{\overbar{SL}}, \smallmath{\underline{SsL}}) plus any additional model to  be included in the evaluation. All the resulting models will be developed on the same labelling budget $\mathcal{L}$. 
    
    \item $(n,k)$ is a pair of integers that regulate the `runs' of CEF-SsL to achieve statistically significant results.
\end{itemize}
The \textit{output} are two ($n\!\!\cdot\!\!k$)-dimensional arrays, whose elements include the results of all the models devised through {\footnotesize$\overrightarrow{\text{ML}}$} on each run, i.e.: {\small$\overrightarrow{\mu}$}, containing the performance on `future' data; and {\small$\overrightarrow{\varepsilon}$}, containing any cost incurred during the development (not related to labelling). 

\begin{figure}[!htbp]
    \centering
    \includegraphics[width=1\columnwidth]{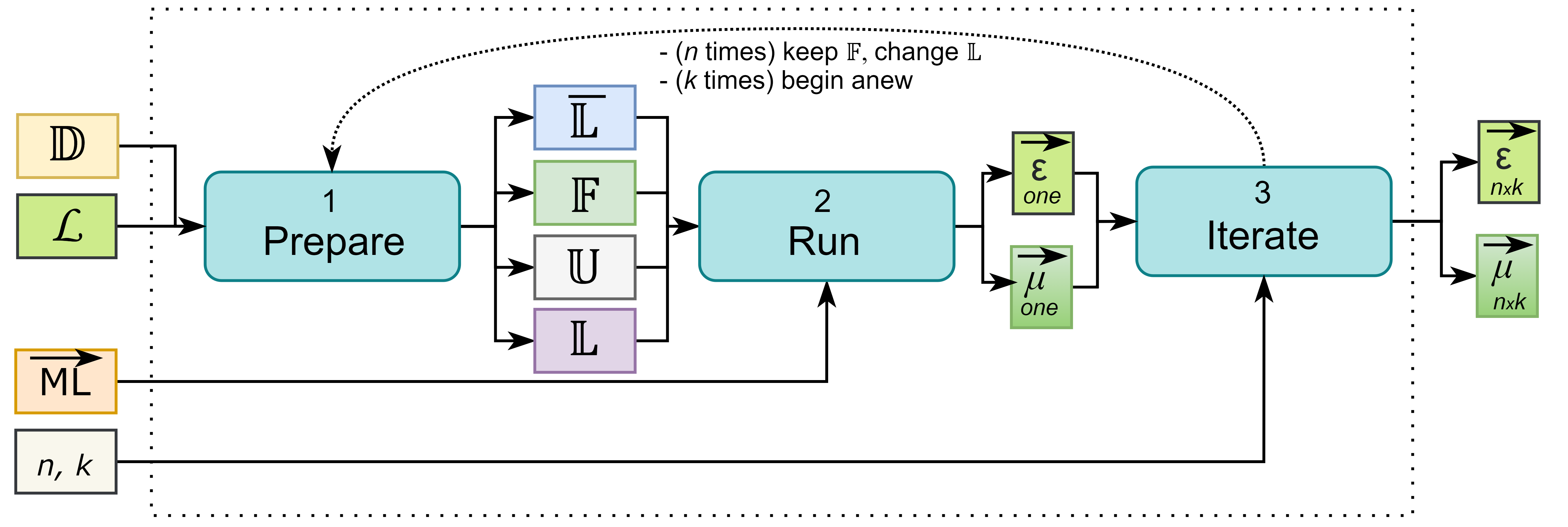}
    \caption{CEF-SsL. $\mathcal{L}$ can also be provided as input for the second stage.}
    \label{fig:overview}
\end{figure}

We provide an overview of CEF-SsL in Fig.~\ref{fig:overview}. CEF-SsL can be divided in three stages: Prepare, Run, Iterate.

\subsection{Stage one: Prepare}
\label{ssec:fmw_first}

\begin{figure}[!htbp]
    \centering
    \includegraphics[width=0.9\columnwidth]{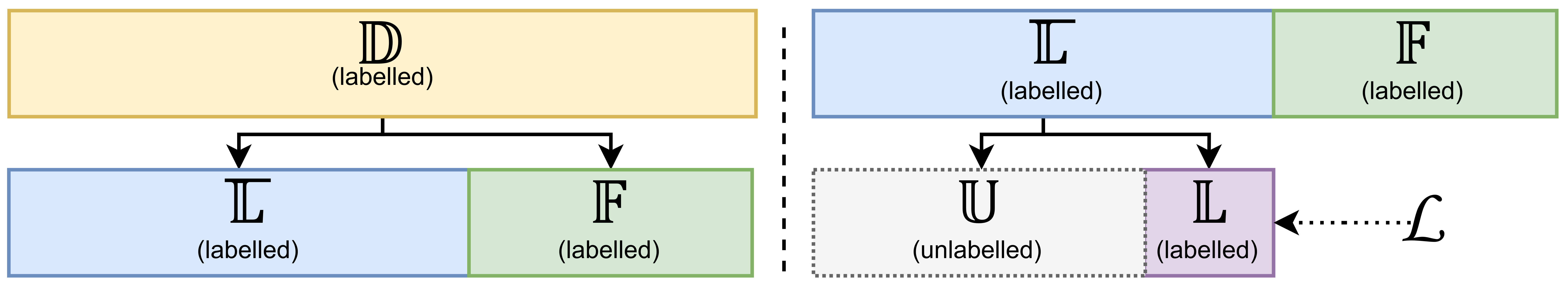}
    \caption{$\mathbb{D}$ is first split into $\mathbb{F}$ and $\mathbb{\overbar{L}}$. Then $\mathbb{\overbar{L}}$ is further split into $\mathbb{L}$ according to $\mathcal{L}$, and the leftout samples are considered as unlabelled $\mathbb{U}$.}
    \label{fig:split}
\end{figure}

The first stage uses $\mathcal{L}$ to partition $\mathbb{D}$ into $\mathbb{L}$, $\mathbb{\overbar{L}}$, $\mathbb{U}$, and $\mathbb{F}$. Fig.~\ref{fig:split} shows a schematic of such a workflow. 

CEF-SsL begins by splitting $\mathbb{D}$ in $\mathbb{F}$ and $\mathbb{\overbar{L}}$: the former, $\mathbb{F}$, is used exclusively to assess the performance on future data\footnote{$\mathbb{F}$ does on depend on the labelling budget $\mathcal{L}$ and $\mathbb{F}\!\cap\!(\mathbb{U}\cup\mathbb{L})$=$\varnothing$.}; the latter, $\mathbb{\overbar{L}}$ is used for all remaining `training' operations, because $\mathbb{\overbar{L}}$ can serve as basis to generate $\mathbb{L}$, and then treat the remaining samples as unlabelled, representing $\mathbb{U}$.

Generating $\mathbb{F}$ from $\mathbb{D}$ depends on the considered CTD task. The selection is done so as to achieve a representative $|\mathbb{F}|$ and $\rho(\mathbb{F})$, while ensuring that the left-out samples (which will represent $\mathbb{\overbar{L}}$) allow to create meaningful $\mathbb{L}$ and $\mathbb{U}$. Such selection can also take into account the temporal relationships (if available) among the samples in $\mathbb{D}$. For example, $\mathbb{F}$ can be composed by selecting only the `most recent' samples in $\mathbb{D}$, allowing assessment of potential concept drift~\cite{jordaney2017transcend}.

To generate $\mathbb{L}$ from $\mathbb{\overbar{L}}$, we recall that $\mathcal{L} \! = \! \sum_{x \in \mathbb{L}} \mathcal{C}_x$. Therefore, CEF-SsL chooses samples from $\mathbb{\overbar{L}}$ and assigns them to $\mathbb{L}$, each time by decreasing the labelling budget $\mathcal{L}$ according to the cost of each sample\footnote{Such cost can be fixed, or can vary depending on the desired level of realistic fidelity by associating each sample $x\!\in\!\mathbb{D}$ with a custom $\mathcal{C}_x$.}. However, $\mathbb{L}$ must include at least some benign and malicious samples. To this purpose, CEF-SsL requires a \textit{minimum} amount of samples for each class. CEF-SsL will then populate $\mathbb{L}$ by randomly sampling as many samples of that class from $\mathbb{\overbar{L}}$ (while decreasing the labelling budget accordingly); then, CEF-SsL will use the remaining budget to further populate $\mathbb{L}$ randomly. We observe that by setting different labelling costs $\mathcal{C}_x$ it is possible to simulate imbalanced data distributions: we will showcase such intriguing property of CEF-SsL in our demonstration.
At this point, CEF-SsL will consider all the samples in $\mathbb{\overbar{L}}$ not included in $\mathbb{L}$ as unlabelled, representing $\mathbb{U}$. The case where one (or many) SsL approaches in {\footnotesize$\overrightarrow{\text{ML}}$} involve the usage of exploratory techniques is covered in the following stage.

\subsection{Stage two: Run}
\label{ssec:fmw_second}
The second stage begins with Reqs.~\ref{req:floor}--\ref{req:ceiling}. Specifically:
\begin{enumerate}
    \item train \smallmath{SL} on $\mathbb{L}$ and test it on $\mathbb{F}$ as \smallmath{\mu(SL)}; account for all operational costs as \smallmath{\varepsilon(SL)};
    \item train \smallmath{\overbar{SL}} on $\mathbb{\overbar{L}}$ and test it on $\mathbb{F}$ as \smallmath{\mu(\overbar{SL})}; account for all operational costs as \smallmath{\varepsilon(\overbar{SL})}.
    \item use $\mathbb{L}$ and $\mathbb{U}$ to devise \smallmath{\underline{SsL}}, and test it on $\mathbb{F}$ as \smallmath{\mu(\underline{SsL})}; account for all operational costs as \smallmath{\varepsilon(\underline{SsL})}.
\end{enumerate}
Then, CEF-SsL focuses on each remaining method in {\footnotesize$\overrightarrow{\text{ML}}$}:
\begin{itemize}
    \item if the SsL method does not make any assumptions on $\mathbb{L}$, then CEF-SsL uses $\mathbb{U}$ and the previously drawn $\mathbb{L}$ as input for the SsL method.
    \item if the SsL method requires $\mathbb{L}$ to be composed in a more refined way, then CEF-SsL generates a new $\mathbb{L}$ according to the specifics of the SsL method. The cost of labelling is `charged' to $\mathcal{L}$ and all other costs are accounted as \smallmath{\varepsilon(SsL)}. The resulting $\mathbb{L}$ and (new) $\mathbb{U}$ are then used as input to the SsL method.
    \item Finally, the respective \smallmath{SsL} model is trained, and tested on $\mathbb{F}$ resulting in \smallmath{\mu(SsL)}; all the operational costs are accounted in \smallmath{\varepsilon(SsL)}. 
\end{itemize}
At the end of this stage, CEF-SsL populates the respective {\small$\overrightarrow{\mu}$} and {\small$\overrightarrow{\varepsilon}$} with the specific performance and (extra) costs of all the considered models of this single `run'.

\subsection{Stage three: Iterate}
\label{ssec:fmw_third}
To obtain statistically significant results, CEF-SsL performs multiple `runs', according to the iteration parameters ($n,k$) provided as input. Specifically:
\begin{itemize}
    \item CEF-SsL repeats its entire workflow $k$ times, each time by choosing a new $\mathbb{F}$, leading to different $\mathbb{\overbar{L}}$ and, hence, different $\mathbb{L}$ and $\mathbb{U}$. 
    \item For each (new) $\mathbb{F}$, CEF-SsL composes a different $\mathbb{L}$ (and, hence, different $\mathbb{U}$) for $n$ times to account for randomness.
\end{itemize}
Altogether, CEF-SsL will evaluate \textit{each method} in {\footnotesize$\overrightarrow{\text{ML}}$} a total of $n \! \cdot \! k$ times, resulting in as many $\mu$ and $\varepsilon$, all of which will be inserted into {\small$\overrightarrow{\mu}$} and {\small$\overrightarrow{\varepsilon}$}.
Assessment of different $\mathcal{L}$ requires CEF-SsL to be restarted by providing a differen $\mathcal{F}$ while maintaining all other inputs.

Finally, the aggregated results of each method can be validated via, e.g., a Student t-test~\cite{zimmerman1987comparative} or a more refined Wilcoxon Ranksum test~\cite{datta2005rank}---provided that CEF-SsL is run enough times to provide statistically significant results (e.g., 50s or more~\cite{happ2019optimal}). Such comparisons can be made  by considering different initial conditions (e.g., different $\mathcal{L}$), because CEF-SsL ensures that all such conditions are shared across all methods: hence, the only difference is which ML method produced each single result.

\section{Demonstration}
\label{sec:demonstration}

As a final contribution of our paper, we demonstrate the application of our CEF-SsL framework to assess the benefits of SsL in CTD. We do so through a massive experimental campaign using 9 well-known datasets, in which we consider 9 existing SsL methods. We aim to:
\begin{itemize}
    \item show how to apply our cost model and CEF-SsL;
    \item further motivate the importance of our requirements;
    \item provide the first statistically validated \textit{benchmark} for future studies.
\end{itemize}
All our considered datasets are publicly available, and we release the code of our CEF-SsL implementation\footnote{Available at: \url{https://github.com/hihey54/CEF-SsL}}.

We first outline the chosen datasets (§\ref{ssec:cs_datasets}). Next, we explain the considered SsL methods (§\ref{ssec:approaches}) and their implementation through CEF-SsL (§\ref{ssec:implementation}). We then summarize the results (§\ref{ssec:evaluation}) and showcase a statistically significant comparison (§\ref{ssec:validation}). We report in the Appendix some low-level details (§\ref{app:testbed}) and the full benchmark results (§\ref{app:benchmark}).

\subsection{Datasets}
\label{ssec:cs_datasets}
Our evaluation focuses on the three CTD areas considered in this paper (§\ref{sec:background}): NID, PWD, MD. For each area we consider three publicly available datasets.

For NID, we use: \dataset{CTU13}, \dataset{UNB15}, \dataset{IDS17}. These datasets are well-known in the NID community, and contain data representing a mixture of synthetic and real network traffic captured in large environments. \dataset{CTU13} is provided as PCAP traces and is focused on \textit{botnet}; \dataset{UNB15} and \dataset{IDS17} are provided as NetFlows and contain additional malicious activities such as DoS, exploits, or illegitimate recon.

For PWD, we use: \dataset{UCI}, \dataset{$\delta$Phish}, \dataset{Mendeley}. These well-known datasets contain webpage information, such as the URL, the reputation of the website, and the contents of the source HTML. Two (\dataset{UCI} and \dataset{Mendeley}) are provided directly as features, while \dataset{$\delta$Phish} has raw webpages, from which we extract the features by following established practices~\cite{mohammad2014intelligent}.

For MD, we use: \dataset{Drebin}, \dataset{Ember}, \dataset{AndMal20}. These datasets are widely employed for ML-related analyses on malware targeting different OS: \dataset{Ember} for Windows, \dataset{Drebin} and \dataset{AndMal20} for Android. Although \dataset{Drebin} is becoming outdated (it was collected in 2013), \dataset{AndMal20} is very recent and serves for a better representation of current trends. 

After obtaining all these 9 datasets, we preprocess them so that they are usable for our objectives. For instance, we clean some redundant data, or derive their feature sets. 
Some datasets may contain samples of different malicious classes. Because our focus is on \textit{detection}, in some cases (\dataset{Drebin}, \dataset{AndMal20}) we aggregate all of them into a single malicious class. Datasets for PWD and \dataset{Ember} only have one malicious class. Datasets for NID have a huge variety: as done in previous work~\cite{apruzzese2020deep}, we consider one classifier per each specific attack. More details are in the Appendix~\ref{sapp:preprocessing}. Regardless, all these operations are fixed for all the ML models evaluated on each dataset, meaning that their impact on $\varepsilon$ is the same for each model and, hence, negligible for comparisons.

We provide an overview of these datasets after all preprocessing has taken place in Table~\ref{tab:datasets}. For each dataset $\mathbb{D}$, we report the overall amount of benign and malicious samples ($\mathbb{D}_b$ and $\mathbb{D}_m$, respectively), the amount of malicious classes (N), the reiterations ($n,k$) performed by CEF-SsL, and reference to a past work that used such dataset (SotA).

\begin{table}[!htbp]
    \centering
    \caption{Considered Datasets}
    \label{tab:datasets}
    \resizebox{0.9\columnwidth}{!}{
        \begin{tabular}{c?c|c|c|c|c|c}
             \textbf{CTD} & \textbf{Name} & $\mathbb{D}_b$ & $\mathbb{D}_m$ & N & ($n,k$) & SotA \\ 
             
             \toprule
             
             \multirow{3}{*}{{\textbf{NID}}} 
              & \dataset{CTU13}~\cite{Garcia:CTU} & 19.5M & 444K & 6 & (11,3) & \cite{Apruzzese:Evaluating} \\
              & \dataset{UNB15}~\cite{UNSWNB15:Dataset} & 100K & 2.22M & 4 & (23,4) & \cite{rajagopal2020stacking}  \\
              & \dataset{IDS17}~\cite{CICIDS2018:Dataset} & 555K & 2.21M & 5 & (15,3) & \cite{di2020experimental}  \\
              
              \midrule
             
              \multirow{3}{*}{{\textbf{PWD}}} 
              & \dataset{UCI}~\cite{karabatak2018performance} & 4898 & 6157 & 1 & (20,5) & \cite{sharma2020feature}  \\
              & \dataset{Mendeley}~\cite{mendeley2020dataset} & 58K & 31K & 1 & (20,5) & \cite{al2021optimized}  \\
              & \dataset{$\delta$Phish}~\cite{Corona:Deltaphish} & 5510 & 1013 & 1 & (20,5) & \cite{Corona:Deltaphish}  \\
              
              \midrule
              
              \multirow{3}{*}{\textbf{MD}}
              
              & \dataset{Drebin}~\cite{arp2014drebin} & 123K & 4022 & 1 & (20,5) & \cite{rana2018evaluating}  \\
              & \dataset{Ember}~\cite{anderson2018ember} & 400K & 400K & 1 & (20,5) & \cite{galen2021empirical}  \\
              & \dataset{AndMal20}~\cite{rahali2020didroid} & 162K & 195K & 1 & (20,5) & \cite{keyes2021entroplyzer}  \\
            
            \bottomrule
        \end{tabular}

    }
\end{table}

For each dataset, CEF-SsL performs all experiments N times, totalling N$*(n\!*\!k)$ runs.

\subsection{Selected SsL Methods}
\label{ssec:approaches}
Let us describe the SsL methods that are included in {\footnotesize$\overrightarrow{ML}$} in out study, together with the two SL baselines. Evaluation of all SsL methods proposed in the state-of-the-art (cf. Table~\ref{tab:sota}) is clearly infeasible and also inpossible due to their limited reproducibility and different assumptions. To showcase all scenarios envisioned in our CEF-SsL framework, we consider 9 SsL methods which are variations of two established SsL methods: \textit{self learning via pseudo-labelling} (e.g.,~\cite{zhang2020semi}) and \textit{active learning via uncertainty sampling} (e.g.,~\cite{rashidi2017android}), summarized in §\ref{ssec:focus}. Specifically, we consider 3 `pure' pseudo-labelling methods, 3 `pure' active learning methods, and 3 combinations thereof (e.g.,~\cite{zhang2021network}), where we cascade pseudo-labelling with active learning. The decision criterion is the \textit{confidence} threshold {\footnotesize$c$}.

\textbf{Pseudo Labelling.} 
One of the `pure' pseudo labelling models represents our \underline{SsL} baseline. We devise:
\begin{itemize}
    \item \smallmath{\underline{SsL}}, using all pseudo labels regardless of their confidence; the process is entirely automated because $\mathbb{L}$ is chosen randomly and no selection of $c$ is required.
    \item \smallmath{\pi\,SsL}, using only the pseudo labels with the highest confidence {\footnotesize$c\geq99\%$};
    \item \smallmath{\widehat{\pi}\,SsL}, which repeats the previous operation another time. We use \smallmath{\pi\,SsL} to predict the remaining $\mathbb{U}$, and insert the corresponding pseudo-labelled samples with {\footnotesize$c\geq99\%$} in the `mixed' $\mathbb{L}$.
\end{itemize}

\textbf{Active Learning.} 
For these methods, we assume that half of the labelling budget is used to develop the first learner, and then the remaining half is used to assign the correct label (specified in the source dataset) to those samples that meet a specific confidence threshold, {\footnotesize$c$}. 
To avoid bias, the `actively labelled' samples are chosen randomly among those that meet the criteria. Due to such randomness, we repeat the draw 5 times for each active learning method.
Depending on {\footnotesize$c$}, we devise:
\begin{itemize}
    \item \smallmath{\alpha\,SsL_l}, using low confidence samples, {\footnotesize$c\leq1\%$};
    \item \smallmath{\alpha\,SsL_h} using high confidence samples {\footnotesize$c \! \geq \! 99\%$};
    \item \smallmath{\alpha\,SsL_o} using the other samples {\footnotesize$1\% < c <99\%$};
\end{itemize}
We note that our implementation of active learning is fundamentally different and more realistic than the one adopted by Pendlebury et al.~\cite{pendlebury2019tesseract}: in Tesseract, the oracle assigns the correct label to \textit{all} samples within a certain confidence; on the other hand, we simply use the learner to provide a set of samples to the oracle, who can only label as many samples as allowed by the remaining budget. This ensures that the provided $\mathcal{L}$ is \textit{never} exceeded, allowing for fair comparisons. 

\textbf{Pseudo-Active Learning.}
We combine pseudo labelling with active learning by using \smallmath{\pi\,SsL} as initial learner (but developed with half of the initial $\mathcal{L}$), which produces three `pseudo-active' methods (\smallmath{\alpha^{\pi}\!SsL_l}, \smallmath{\alpha^{\pi}\!SsL_o}, \smallmath{\alpha^\pi\!SsL_h}) in the same way as in the `pure' active learning.

\subsection{Implementation}
\label{ssec:implementation}
In our implementation, CEF-SsL performs the same operations for all considered $\mathbb{D}$ and {\footnotesize$\overrightarrow{ML}$}.

The first stage of CEF-SsL is the creation of $\mathbb{F}$ from $\mathbb{D}$. In our case, we do so by adopting the 80:20 split which is common in CTD (e.g.,~\cite{apruzzese2020deep, alshahrani2018ddefender, awasthi2021phishing}). Specifically, CEF-SsL randomly chooses 20\% of the malicious samples in $\mathbb{D}$ and 20\% of the benign samples in $\mathbb{D}$, and puts them into $\mathbb{F}$. The resulting samples are then considered as $\mathbb{\overbar{L}}$. 

To allow a comprehensive benchmark, we consider three scenarios where the cost of labelling each sample varies depending on their class---which serves to investigate different \textit{balance ratios} $\rho(\mathbb{L})$. 
Specifically:
\begin{itemize}
    \item (balanced) {\small$\mathcal{C}_m \! = \mathcal{C}_b$}: the number of benign samples matches that of malicious, {\small$\rho(\mathbb{L}) \! = \! (50, 50)$};
    \item (unbalanced) {\small$\mathcal{C}_m \!\! = \! 2 \mathcal{C}_b$}: the number of benign samples is twice that of malicious, {\small$\rho(\mathbb{L}) \! = \! (66, 33)$};
    \item (very unbalanced) {\small$\!\mathcal{C}_m\!\!=\!\!5 \mathcal{C}_b$}: the number of benign samples is five times that of malicious, {\small$\rho(\mathbb{L})\!\!=\!\!(84, 16)$}.
\end{itemize}
Where $b$ and $m$ denote a benign and malicious sample, respectively.
For each cost scenario, we vary the allocated labelling budget $\mathcal{L}$ four times. We do so by regulating the minimal amount of \textit{benign} samples, $\mathbb{L}_b$, to be included in $\mathbb{L}$. Hence, CEF-SsL composes $\mathbb{L}$ by first selecting $\mathbb{L}_b$ benign samples from $\mathbb{\overbar{L}}$; then, CEF-SsL keeps populating $\mathbb{L}$ by choosing malicious samples from $\mathbb{\overbar{L}}$ until the budget $\mathcal{L}$ is depleted, resulting in $\mathbb{L}_m$ malicious samples. The values of $\mathcal{L}$ and $\mathbb{L}_b$ depend on each CTD task, and are reported in Table~\ref{tab:scenarios}. Because each combination of $\mathcal{L}$ and $\mathcal{C}$ represents a different setting, we restart CEF-SsL at each change (hence, 12 times) to allow a fair comparison.

\begin{table}[!htbp]
    \centering
    \caption{Composition of $\mathbb{L}$ for different $\mathcal{L}$ and $\mathcal{C}$. In all cases, $|\mathbb{L}|$=($\mathbb{L}_m$+$\mathbb{L}_b$),  $\mathbb{F}$=0.2*$\mathbb{D}$, $\mathbb{\overbar{L}}$=0.8*$\mathbb{D}$, and $\mathbb{U}$=($\mathbb{\overbar{L}}$-$\mathbb{L}$).}
    \label{tab:scenarios}
    \resizebox{0.8\columnwidth}{!}{
    
        \begin{tabular}{c?c|c|c?c|c|c?c|c|c}
             
             CTD & \multicolumn{3}{c?}{\textbf{NID}} & \multicolumn{3}{c?}{\textbf{PWD}} & \multicolumn{3}{c}{\textbf{MD}} \\ \cline{1-10}
             Scenario and $\mathcal{C}$ & $\mathcal{L}$ & $\mathbb{L}_m$ & $\mathbb{L}_b$ & $\mathcal{L}$ & $\mathbb{L}_m$ & $\mathbb{L}_b$ & $\mathcal{L}$ & $\mathbb{L}_m$ & $\mathbb{L}_b$ \\
             
             \midrule
             
             \multirow{4}{*}{
                 \begin{tabular}{c}
                    balanced \\
                    $\mathcal{C}_b \! = \! \mathcal{C}_m$ \\
                 \end{tabular}
             } 
             & 100 & $50$ & $50$ & 40 & $20$ & $20$ & 80 & $40$ & $40$ \\
             & 200 & $100$ & $100$ & 80 & $40$ & $40$ & 160 & $80$ & $80$ \\
             & 400 & $200$ & $200$ & 160 & $80$ & $80$ & 320 & $160$ & $160$ \\
             & 800 & $400$ & $400$ & 320 & $160$ & $160$ & 640 & $320$ & $320$ \\
             
             \midrule
             
             \multirow{4}{*}{
                \begin{tabular}{c}
                    unbalanced \\
                    $\mathcal{C}_m \! = \! 2*\mathcal{C}_b$ \\
                 \end{tabular}
            }
             & 200 & $50$ & $100$ & 80 & $20$ & $40$ & 160 & $40$ & $80$ \\
             & 400 & $100$ & $200$ & 160 & $40$ & $80$ & 320 & $80$ & $160$ \\
             & 800 & $200$ & $400$ & 320 & $80$ & $160$ & 640 & $160$ & $320$ \\
             & 1600 & $400$ & $800$ & 640 & $160$ & $320$ & 1280 & $320$ & $640$ \\
             
             \midrule
             
             \multirow{4}{*}{
             \begin{tabular}{c}
                    very \\
                    unbalanced \\
                    $\mathcal{C}_m \! = \! 5*\mathcal{C}_b$ \\
                 \end{tabular}
             } 
             & 500 & $50$ & $250$ & 200 & $20$ & $100$ & 400 & $40$ & $200$ \\
             & 1000 & $100$ & $500$ & 400 & $40$ & $200$ & 800 & $80$ & $400$ \\
             & 2000 & $200$ & $1000$ & 800 & $80$ & $400$ & 1600 & $160$ & $800$ \\
             & 4000 & $400$ & $2000$ & 1600 & $160$ & $800$ & 3200 & $320$ & $1600$ \\
             
                          \bottomrule
        \end{tabular}

    }
\end{table}

We observe that our choices of $\mathcal{L}$ result in $\mathbb{L}$ that are \textit{smaller} than the testing set $\mathbb{F}$, which is a good practice in CTD research~\cite{marchal2018designing}. Overall, our models are trained with as little as 40 labels (for PWD), and as high as 2400 labels (for NID) 
The resulting sets $\mathbb{\overbar{L}}$ and $\mathbb{L}$ are immediately used to train the lower bound \smallmath{SL} and the upper bound \smallmath{\overbar{SL}}, both tested on $\mathbb{F}$. Then, CEF-SsL uses the remaining sets according to each SsL method in {\footnotesize$\overrightarrow{ML}$}. More detailed information on such development can be found in Appendix~\ref{app:ssl-methods}.

\subsection{Evaluation}
\label{ssec:evaluation}
We apply the described implementation of CEF-SsL on each dataset.
The considered ML methods use the Random Forest (RF) learning algorithm, for three reasons. First, because RF are widely adopted by past work (e.g.,~\cite{apruzzese2020deep,rana2018evaluating,galen2021empirical, pendlebury2019tesseract}), favoring comparisons. Second, since RF are known to provide an excellent tradeoff between performance and computation time, and they can also be parallelized: to provide statistically significant results we must consider thousands of models, hence we favor algorithms that are fast to train. Third, because preliminary analyses confirmed the previous statement: we empirically found that RF achieve similar performance as other algorithms (e.g., neural networks) while requiring a fraction of the training time.\footnote{We do not aim at benchmarking every conceivable implementation of SsL methods. Nevertheless, our CEF-SsL code allows to select a different learning algorithm by changing just one line of code.}

\textbf{Performance Assessment.}
We choose the \textit{F1-score} (a positive is a malicious sample) as performance metric, which is common in CTD.\footnote{We also measure Recall and Precision, reported in our GitHub.} We report in Table~\ref{tab:fscores_avg} the \textit{average} F1-score achieved by each method in \smallmath{\overrightarrow{ML}} on each dataset, across all the different combinations of $\mathcal{L}$ and $\mathcal{C}$. Granular analyses and model-to-model comparisons can be made by looking at the full results in Appendix~\ref{app:benchmark}.

\begin{table}[!htbp]
    \centering
    \caption{Average F1-score of all methods. Boldface denotes the best SsL method on each {\footnotesize{$\mathbb{D}$}}. Gray cells denote the best `pure' pseudo labelling method, while dark-gray is for the best active learning method.}
    \label{tab:fscores_avg}
    \resizebox{\columnwidth}{!}{
    
        \begin{tabular}{c?c|c|c?c|c|c?c|c|c|}
             
             \multicolumn{1}{c?}{\textbf{CTD}}&
             \multicolumn{3}{c?}{\textbf{NID}} & 
             \multicolumn{3}{c?}{\textbf{PWD}} & 
             \multicolumn{3}{c|}{\textbf{MD}} \\ 
             \cline{1-10}
             
             \multicolumn{1}{c?}{Method} & \dataset{CTU13} & \dataset{UNB15} &   \dataset{IDS17} & \dataset{Mend} & \dataset{UCI} & \dataset{$\delta$Phish} & \dataset{DREBIN} & \dataset{Ember} & \dataset{AndMal} \\
             
             \toprule
             
             
             \smallmath{\overbar{SL}} & $0.979$ & $0.942$ & $0.989$ & $0.958$ & $0.974$ & $0.958$ & $0.907$ & $0.970$ & $0.986$ \\
             
             \smallmath{SL} & $0.611$ & $0.447$ & $0.878$ & $0.852$ & $0.884$ & $0.780$ & $0.480$ & $0.667$ & $0.910$\\
             
             \smallmath{\underline{SsL}} & \cellcolor{gray!20}$0.613$ & \cellcolor{gray!20}$0.447$ & \cellcolor{gray!20}$0.879$ & \cellcolor{gray!20}$0.852$ & \cellcolor{gray!20}$0.886$ & \cellcolor{gray!20}$\mathbf{0.778}$ & \cellcolor{gray!20}$0.486$ & \cellcolor{gray!20}$0.662$ & \cellcolor{gray!20}$0.910$\\
             
             \midrule
             
             \smallmath{\pi\,SsL} & $0.588$ & $0.437$ & $0.820$ & $0.850$ & $0.884$ & $0.778$ & $0.474$ & $0.647$ & $0.900$\\
             
             \smallmath{\widehat{\pi}\,SsL} & $0.584$ & $0.435$ & $0.818$ & $0.849$ & $0.883$ & $0.777$ & $0.470$ & $0.641$ & $0.890$ \\
             
             \midrule
             
             \smallmath{\alpha\,SsL_l} & \cellcolor{gray!40}$\mathbf{0.693}$ & $0.582$ & \cellcolor{gray!40}$\mathbf{0.897}$ & \cellcolor{gray!40}$\mathbf{0.863}$ & \cellcolor{gray!40}$\mathbf{0.903}$ & \cellcolor{gray!40}$0.770$ & \cellcolor{gray!40}$\mathbf{0.546}$ & \cellcolor{gray!40}$\mathbf{0.687}$ & \cellcolor{gray!40}$\mathbf{0.924}$\\
             
             \smallmath{\alpha\,SsL_o} & $0.637$ & $0.577$ & $0.874$ & $0.855$ & $0.891$ & $0.745$ & $0.497$ & $0.673$ & $0.916$ \\
             
             \smallmath{\alpha\,SsL_h} & $0.510$ & $0.436$ & $0.786$ & $0.834$ & $0.851$ & $0.714$ & $0.423$ & $0.598$ & $0.892$ \\
             
             \midrule
             \smallmath{\alpha^{\pi}\!SsL_l} & $0.664$ & $0.533$ & $0.853$ & $0.861$ & $0.901$ & $0.767$ & $0.529$ & $0.654$ & $0.901$ \\
             
             \smallmath{\alpha^{\pi}\!SsL_o} & $0.633$ & \cellcolor{gray!40}$\mathbf{0.595}$ & $0.857$ & $0.854$ & $0.890$ & $0.745$ & $0.489$ & $0.647$ & $0.895$ \\
             
             \smallmath{\alpha^{\pi}\!SsL_h} & $0.486$ & $0.427$ & $0.744$ & $0.833$ & $0.851$ & $0.711$ & $0.410$ & $0.579$ & $0.865$ \\

             \bottomrule
        \end{tabular}
    }
\end{table}

The following insights can be drawn from Table~\ref{tab:fscores_avg}.
First, albeit almost counter-intuitive, using \textit{all} pseudo-labels is the most effective among the `pure' pseudo labelling techniques.
Second, despite the identical labelling budgets in NID datasets, SsL methods achieved varying performance: in \dataset{UNB15} they achieve only 0.6 F1-score at best, whereas in \dataset{IDS17} they can reach almost 0.9 F1-score; this highlights the importance of conducting evaluations on diverse datasets. 
Third, active learning appears to be the best way to use the labelling budget, but in \dataset{$\delta$Phish} it is always inferior to `pure' pseudo labelling; moreover, it is interesting that the results of the models trained on the (correct) high confidence labels consistently achieve the worst performance.
Fourth, in all datasets, all models performed very similarly (on average) to the baseline \smallmath{SL}. Such small gap requires to be further investigated via statistical comparisons, which we will do in §\ref{ssec:validation}.

\textbf{Assessment of Extra Costs.}
Many factors contribute to $\varepsilon$ for each considered ML method. All of our implemented methods share the same testbed, and most of such costs are equal for all methods. We hence focus on the most salient `cost' of each method that we can measure: its \textit{execution time}. 

We report in Table~\ref{tab:times_avg} the total time required to develop each model, which comprises all the steps for (re)training and (for SsL methods) predicting unlabelled data.

\begin{table}[!htbp]
    \centering
    \caption{Average execution time (seconds) of all methods. Bold values denote the SsL method with best F1-score on the same {\footnotesize$\mathbb{D}$} (cf. Table~\ref{tab:fscores_avg}).}
    \label{tab:times_avg}
    \resizebox{\columnwidth}{!}{
    
        \begin{tabular}{c?c|c|c?c|c|c?c|c|c|}
             
             \multicolumn{1}{c?}{\textbf{CTD}}&
             \multicolumn{3}{c?}{\textbf{NID}} & 
             \multicolumn{3}{c?}{\textbf{PWD}} & 
             \multicolumn{3}{c|}{\textbf{MD}} \\ 
             \cline{1-10}
             
             \multicolumn{1}{c?}{Method} & \dataset{CTU13} & \dataset{UNB15} &       \dataset{IDS17} & \dataset{Mend} & \dataset{UCI} & \dataset{$\delta$Phish} & \dataset{DREBIN} & \dataset{Ember} & \dataset{AndMal} \\
             
             \toprule
             
             
             $\overbar{SL}$ & $30.31$ & $18.75$ & $33.55$ & $1.365$ & $0.420$ & $0.535$ & $3.054$ & $147.6$ & $42.81$\\
             
             $SL$ & $0.392$ & $0.388$ & $0.393$ & $0.349$ & $0.390$ & $0.401$ & $0.381$ & $0.395$ & $0.438$ \\
             $\underline{SsL}$ & $35.00$ & $24.44$ & $39.65$ & $1.199$ & $1.036$ & $\mathbf{1.040}$ & $1.430$ & $101.4$ & $30.53$ \\
             
             \midrule
             
             $\pi\,SsL$ & $12.74$ & $15.32$ & $23.92$ & $1.090$ & $0.930$ & $0.942$ & $1.064$ & $2.257$ & $3.702$ \\
             $\widehat{\pi}\,SsL$ & $27.00$ & $28.77$ & $45.13$ & $1.864$ & $1.473$ & $1.487$ & $1.824$ & $6.791$ & $8.726$ \\

\midrule

$\alpha\,SsL_l$ & $\mathbf{3.471}$ & $4.955$ & $\mathbf{8.990}$ & $\mathbf{0.905}$ & $\mathbf{0.885}$ & $0.895$ & $\mathbf{0.847}$ & $\mathbf{0.897}$ & $\mathbf{0.960}$ \\

$\alpha\,SsL_o$ & $3.469$ & $4.954$ & $8.989$ & $0.904$ & $0.883$ & $0.896$ & $0.846$ & $0.895$ & $0.957$ \\

$\alpha\,SsL_h$ & $3.466$ & $4.950$ & $8.987$ & $0.898$ & $0.880$ & $0.894$ & $0.844$ & $0.893$ & $0.952$ \\

\midrule

$\alpha^{\pi}\!SsL_l$ & $23.49$ & $27.22$ & $38.50$ & $1.744$ & $1.356$ & $1.375$ & $1.666$ & $5.267$ & $7.593$ \\

$\alpha^{\pi}\!SsL_o$ & $23.39$ & $\mathbf{26.98}$ & $38.48$ & $1.746$ & $1.354$ & $1.375$ & $1.662$ & $5.493$ & $7.699$ \\

$\alpha^{\pi}\!SsL_h$ & $23.28$ & $26.78$ & $38.06$ & $1.747$ & $1.350$ & $1.372$ & $1.655$ & $5.258$ & $7.579$ \\
             \bottomrule
        \end{tabular}
    }
\end{table}

From Table~\ref{tab:times_avg}, we can observe that the model requiring the highest time is often the baseline \smallmath{\underline{SsL}}. This is not surprising, because it is trained on the \textit{entire} $\mathbb{U}$ after training the baseline \smallmath{SL}, and using it to predict the entire $\mathbb{U}$. In contrast, all other models are trained on a much smaller dataset.

However, using only the \textit{execution} time when comparing the $\varepsilon$ is not always fair. Some models may be better in terms of execution time but require some manual tuning, e.g., setting the desired level of confidence $c$. Compare, for instance, \smallmath{\underline{SsL}} and \smallmath{\alpha\,SsL_l} on \dataset{CTU13}: the former requires 35s, whereas the latter only 3s, which is a 32s difference. However, choosing an appropriate $c$ for \smallmath{\alpha\,SsL_l} requires: (i) inspecting the results of \smallmath{\alpha\,SsL_l}, (ii) setting the new $c$, (iii) devising a new \smallmath{\alpha\,SsL_l}, (iv) inspecting the results of the new \smallmath{\alpha\,SsL_l}, and (v) deciding whether it has acceptable performance or not. These procedures have a human in the loop and hence a significantly higher $\varepsilon$ (which cannot be shown in Table~\ref{tab:times_avg}). On the other hand, \smallmath{\underline{SsL}} \textit{always} achieves the reported performance, and by being entirely automated will result in an overall lower $\varepsilon$.

\subsection{Statistical Validation}
\label{ssec:validation}
To substantiate the claim that some SsL using $\mathbb{U}$ outperforms the respective baseline (\smallmath{SL} or \smallmath{\underline{SsL}}), statistical tests can be carried out. Here, we use the Wilcoxon ranksum~\cite{datta2005rank}, in which two populations are compared with the goal of verifying a given null hypothesis $H_0$. The test outputs a $z$-value used to derive a $p$-value: $H_0$ can be accepted or rejected on the basis of $p$, according to a target significance level. 
We use the two-tailed version of the test, hence $H_0$ is that the two populations are statistically equivalent: the larger the $p$-value, the more $H_0$ should be accepted (and viceversa). We set the significance level to $0.05$, implying that if $p\!>\!0.05$ then the two populations are equivalent; conversely, if $p\!\leq\!0.05$, the two populations are different (this is especially true if $p\!\ll\!0.05$).

For each dataset, we compare the populations containing the performance of the baseline \smallmath{SL} against: (i) the best `pure' pseudo-labelling method (gray cells in Table~\ref{tab:fscores_avg}); and (ii) the best active learning method (dark gray cells in Table~\ref{tab:fscores_avg}). We are comparing\footnote{The test is valid: the compared populations have the same amount of elements and the conditions are shared among all elements, where the only difference is the generation process (i.e., the specific ML method).} the \textit{methods}, not the individual \textit{models} (model-to-model comparisons can be made from Figs.~\ref{fig:results_nid}--\ref{fig:results_MD}).
The results of such tests are in Table~\ref{tab:statistical}, reporting the size of the populations\footnote{Such size is given by $n$*$k$*$12$, because we consider 3 cost scenarios with 4 budgets. For NID, each element is the average of the N malicious classes.}, and the output $z$- and $p$ values.\footnote{The \textit{EffectSize} of the test can be dervied by $\frac{z}{\sqrt{\text{PopSize}}}$.}



\newcommand{\inc}{\cellcolor{green!15}}
\newcommand{\dec}{\cellcolor{red!15}}

\begin{table}[!htbp]
    \centering
    \caption{We statistically compare the SL baseline method against the best `pure' pseudo-labelling and the best active learning methods. Bold values denote when $H_0$ is accepted ({\scriptsize$p\!>\!0.05$}), i.e., the two methods are statistically equivalent. Cells in green (red) denote cases where using $\mathbb{U}$ statistically increases (decreases) performance.}
    \label{tab:statistical}
    \resizebox{0.95\columnwidth}{!}{
        \begin{tabular}{c|c?c|c|c?c|c|c}
            \multicolumn{2}{c?}{} & \multicolumn{3}{c?}{Best `pure' pseudo-labelling} & \multicolumn{3}{c}{Best active learning} 
            \\ \cline{3-8}
             \textbf{Dataset} & PopSize
             & Method & $p$-value & $z$-value
             & Method & $p$-value & $z$-value \\ 
             
             \toprule

               \dataset{CTU13} & 396 & \smallmath{\underline{SsL}} & $\mathbf{0.873}$ & $0.159$ & \inc{} \smallmath{\alpha\,SsL_l} & \inc{} $<0.001$ & \inc{} $4.310$ \cellcolor{green!15} \\
              \dataset{UNB15} & 1104 & \underline{\smallmath{SsL}} & $\mathbf{0.964}$ & $\text{--}0.044$ & \inc{} \smallmath{\alpha^{\pi}\!SsL_o} & \inc{} $<0.001$ & \inc{} $15.98$ \\
              \dataset{IDS17} & 540 & \underline{\smallmath{SsL}} & $\mathbf{0.932}$ & $0.085$ &  \smallmath{\alpha\,SsL_l} & $\mathbf{0.978}$ & $\text{--}0.027$ \\
              
              \midrule

              \dataset{UCI} & 1200 & \underline{\smallmath{SsL}} & $\mathbf{0.473}$ & $0.717$ & \inc{} \smallmath{\alpha\,SsL_l} & \inc{} $<0.001$ & \inc{} $7.386$ \\
              \dataset{Mend.} & 1200 & \underline{\smallmath{SsL}} & $\mathbf{0.713}$ & $0.368$ & \inc{} \smallmath{\alpha\,SsL_l} & \inc{} $<0.001$ & \inc{} $6.757$ \\
              \dataset{$\delta$Phish} & 1200 & \underline{\smallmath{SsL}} & $\mathbf{0.554}$ & $\text{--}0.590$ & \dec{} \smallmath{\alpha\,SsL_l} & \dec{} $0.002$ & \dec{} $\text{--}3.113$ \\
              
              \midrule

              \dataset{Drebin} & 1200 & \underline{\smallmath{SsL}} & $\mathbf{0.310}$ & $1.015$ & \inc{} \smallmath{\alpha\,SsL_l} & \inc{} $<0.001$ & \inc{} $11.78$ \\
              \dataset{Ember} & 1200 & \underline{\smallmath{SsL}} & $\mathbf{0.603}$ & $\text{--}0.512$ & \inc{} \smallmath{\alpha\,SsL_l} & \inc{} $<0.001$ & \inc{} $3.407$ \\
              \dataset{AndMal} & 1200 & \underline{\smallmath{SsL}} & $\mathbf{0.712}$ & $\text{--}0.370$ & \inc{} \smallmath{\alpha\,SsL_l} & \inc{} $<0.001$ & \inc{} $12.01$ \\
            
            \bottomrule
        \end{tabular}

    }
\end{table}

From Table~\ref{tab:statistical}, we can draw the following conclusions. 

Active Learning provides \textbf{statistically significant improvements}, which can be remarkable (cf. Figs~\ref{fig:results_nid}--\ref{fig:results_MD}); a finding that is consistent with past works (e.g.,~\cite{gornitz2009active}). However, this is not always true: on \dataset{IDS17}, \smallmath{\alpha\,SsL_l} is statistically equivalent to \smallmath{SL}, meaning that there is \textit{no} benefit in using $\mathbb{U}$ on \dataset{IDS17} (at least according to our testbed). Moreover, it can be \textit{detrimental}: on \dataset{$\delta$Phish}, the best method using active learning (\smallmath{\alpha\,SsL_l}) yields lower performance than the baseline \smallmath{SL}, and such difference is statistically significant ($p$=0.002$\ll$0.05).

Pseudo Labelling \textbf{is not useless}. In \dataset{UNB15}, the pseudo-active method \smallmath{\alpha^{\pi}\!SsL_o} statistically outperforms the baselines (\smallmath{SL}, and also \smallmath{\underline{SsL}}). However, in all its `pure' applications, it provides no benefit: its best performer is \smallmath{\underline{SsL}}, which is \textit{always} statistically equivalent to \smallmath{SL}. To put it differently, in the `worst case'  using $\mathbb{U}$ will not induce performance loss.

Finally, we also conducted the one-tailed variant of the statistical test, which confirms all previous findings.

\section{Discussion and Future Work}
\label{sec:discussion}

\noindent
We make some crucial remarks on our evaluation.

\begin{figure*}[!htbp]
     \centering
     \begin{subfigure}[b]{0.9\textwidth}
         \centering
         \includegraphics[width=\columnwidth]{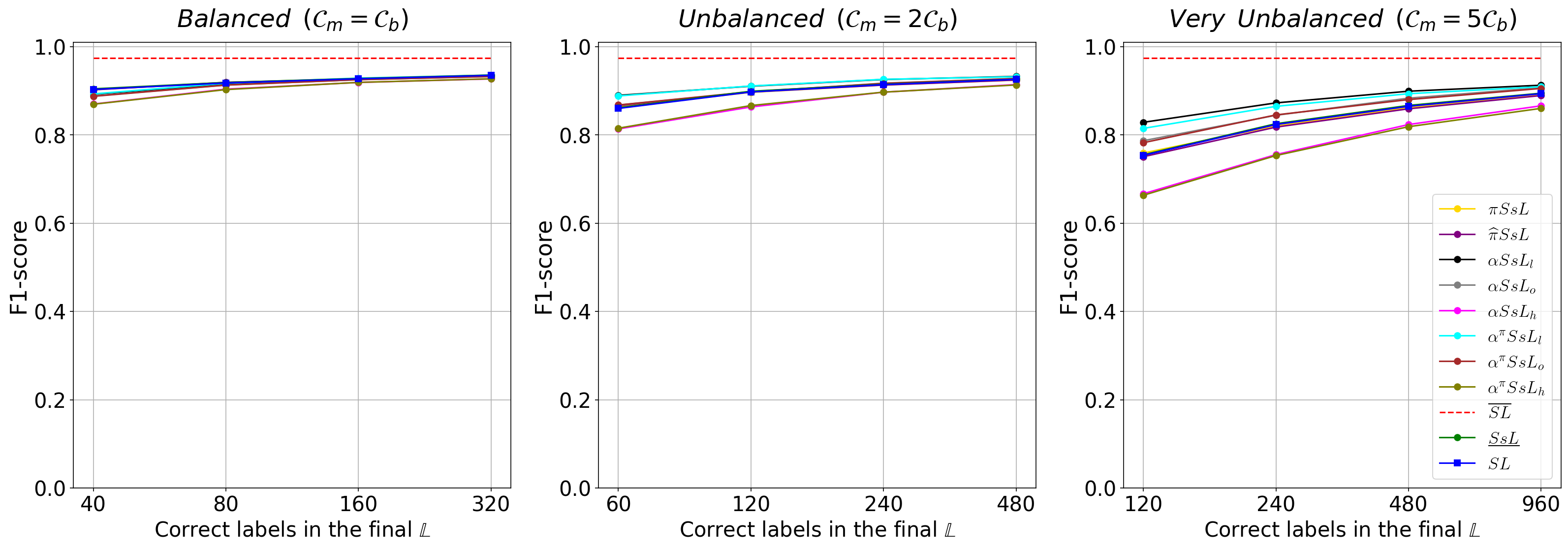}
         \caption{Results on \dataset{\textbf{UCI}}. For every plot, each `point' represents the average results of 100 models (and 5 times as much for all models using active learning).}
         \label{sfig:uci}
     \end{subfigure}
     \begin{subfigure}[b]{0.9\textwidth}
         \centering
         \includegraphics[width=\columnwidth]{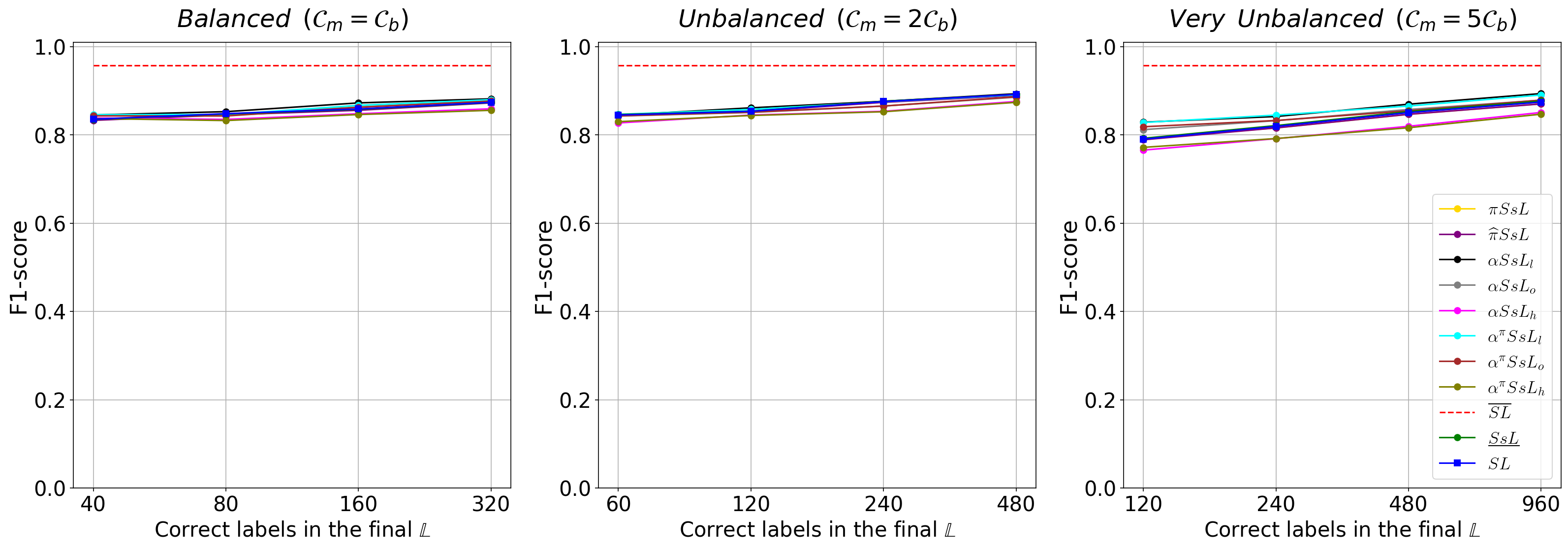}
         \caption{Results on \dataset{\textbf{Mendeley}}. For every plot, each `point' represents the average results of 100 models (and 5 times as much for all models using active learning).}
         \label{sfig:mendeley}
     \end{subfigure}
     \begin{subfigure}[b]{0.9\textwidth}
         \centering
         \includegraphics[width=\columnwidth]{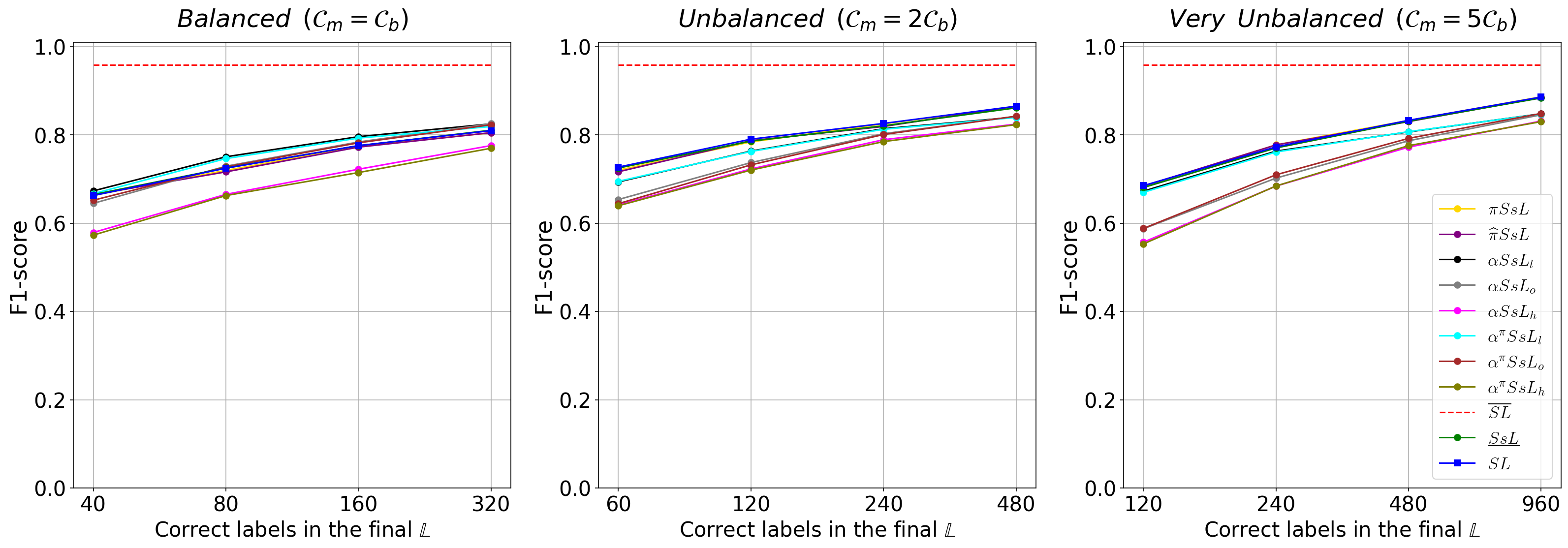}
         \caption{Results on \dataset{$\delta$\textbf{Phish}}. For every plot, each `point' represents the average results of 100 models (and 5 times as much for all models using active learning).}
         \label{sfig:deltaphish}
     \end{subfigure}
     %
     \caption{\textbf{Phishing Website Detection}. For each dataset, we report the results for the three cost (or balancing) scenarios. Each scenario is shown in a plot, where the y-axis reports the F1-score and the x-axis the (increasing) labelling budget. Each method is denoted with a line on each plot.}
    \label{fig:results_pd}
\end{figure*}

\textbf{Importance of our Requirements.}
Without considering \smallmath{SL}, it was not possible to determine that any `pure' pseudo-labelling model was not just useless but even detrimental, due to achieving the same or inferior $\mu(\cdot)$ while requiring higher $\mathcal{B}$. For instance, consider the detailed results on \dataset{UCI} in Fig.~\ref{sfig:uci}: the SsL models achieve a respectable F1-score of 0.9 when trained only on 40 labelled examples. However, the same result is achieved by the baseline \smallmath{SL}, which does not make use of an $\mathbb{U}$ containing 8000 samples. Similarly, on \dataset{AndMal20}, investing in $\mathcal{U}$ provides very little benefit because the performance gap between \smallmath{SL} and \smallmath{\overbar{SL}} is marginal.
Furthermore, active learning was considered to be superior to random sampling~\cite{gu2014active}, but for two of our datasets, \dataset{IDS17} and \dataset{$\delta$Phish}, this cannot be confirmed. Such insights would not have been possible without a massive and statistically validated comparison: looking only at Table~\ref{tab:fscores_avg}, one may conclude that the best active learning model has better average performance than the \smallmath{SL} baseline.
Finally, by considering Req.~\ref{req:random}, it was possible to determine the lowest \textit{cost} induced by using $\mathbb{U}$ in the SsL pipeline due to lack of human supervision (cf. Table~\ref{tab:times_avg}).

\textbf{Balancing.} An intriguing occurrence can be observed from Figs.~\ref{fig:results_nid}--\ref{fig:results_MD}. In the presence of data-imbalance, the performance can be \textit{lower} despite a \textit{higher} labelling budget. This is evident for \dataset{Ember} (Fig.~\ref{sfig:ember}). In the balanced scenario where $\mathcal{C}_m$=$\mathcal{C}_b$ (leftmost plot): when $\mathbb{L}$ contains 800 correct labels, all models converge to a 0.8 F1-score; however, when $\mathcal{C}_m$=5$\mathcal{C}_b$ (rightmost plot), the performance when $\mathbb{L}$ has 2400 correct labels ranges from 0.55 to 0.75 F1-score. We tried to regulate these situations by applying oversampling techniques (e.g.,~\cite{barua2012mwmote}), but we never saw significant changes. Such interesting result also occurs on \dataset{AndMal20} and \dataset{UCI} despite at a lower magnitude.

\textbf{Scope.} 
Despite our extensive evaluation, we stress that our results should serve as a basis for future works\footnote{We will update our GitHub repository with new findings if such findings are derived from CEF-SsL.}, and should not be used to derive `universal' statements. For instance, the considered SsL methods represent just a subset of all conceivable SsL techniques: hence our experiments cannot be used to conclude that ``all pseudo-labelling techniques are not very effective in CTD''. In addition, we stress that we apply well-known methods on existing datasets and do not claim superiority over past works. In Appendix~\ref{app:casestudy}, we present a case study comparing our results with a recent paper~\cite{zhang2021network} using a similar testbed as ours. Furthermore, although our cost model (cf. §\ref{ssec:deployment}) can represent any classification problem, in our evaluation we use CEF-SsL to assess binary classifiers. This is because the main goal of CTD is separating threats from legitimate events, but we acknowledge that some applications may involve fine-grained analyses. Evaluations of SsL methods in multi-classification settings are challenging, and we provide some considerations in Appendix~\ref{app:multiclass}.

\textbf{Labelling Accuracy.} We assume that all the considered datasets are correctly labelled. However, such assumption may be overly optimistic: as described in §\ref{sec:background}, manual labelling is an error-prone task, and some recent papers highlighted that even well-known datasets may contain flaws (e.g.,~\cite{engelen2021troubleshooting}). Due to the seminal nature of this SoK paper, we do not make any change to the provided labels---which also facilitates comparisons with previous works using such datasets, as the ground truth is the same. Nevertheless, we endorse future studies to question the correctness of the labelling procedures---especially if aimed at realistic deployments, as wrong labels induce data poisoning (which can also affect unlabelled samples~\cite{carlini2021poisoning}).

\textbf{Concept Drift.} Our evaluation assumes that the data distribution is stationary. In reality evaluations should be performed at regular intervals to prevent concept drift (cf. §\ref{ssec:deployment}). Such operations, however, are facilitated by the design of our framework: CEF-SsL can also be applied to \textit{lifelong learning} scenarios by selecting $\mathbb{F}$ and $\overbar{\mathbb{L}}$ on a temporal basis. For instance, in Tesseract~\cite{pendlebury2019tesseract}, the initial model is developed under the assumption that all samples from the past are correctly labelled, resulting in over 50K samples. Our CEF-SsL can be applied by assuming that only a fraction of `past' samples are correctly labelled ($\mathbb{L}$) while the remaining ones are not verified ($\mathbb{U}$), and the most recent samples are used as test ($\mathbb{F}$).



\section{Conclusions}
\label{sec:conclusions}
Acquiring ground truth information in CTD is difficult, but large amounts of unlabelled data are regularly available. These premises make SsL an intriguing opportunity, as it exploits unlabelled data to mitigate the problem of scarce ground truth. While many previous works have employed SsL for diverse CTD tasks, \emph{none of them} investigated the benefit provided by unlabelled data. Despite being relatively cheap, such data still brings certain costs into the ML pipeline.

In this SoK paper, we specifically investigate the utility of unlabelled data and hence facilitate deployment of SsL methods for CTD. We formalize the evaluation requirements
that enable one to assess the impact of unlabelled data in the development of a SsL model, under the assumption of a limited labelling budget. Prior works in this area had different scope and never considered such requirements, hence the impact of unlabelled data could not be assessed. 

As a constructive demonstration, we provide an evaluation framework that meets all these requirements and use it to perform the first statistically validated benchmark of 9 selected SsL methods on 9 well-known datasets for CTD.
The results reveal that only SsL methods using active learning are statistically better than baselines that do not use unlabelled data; however, in some cases they can degrade performance.

Our paper hence highlights the substantial margin for improvement of SsL methods for CTD. This motivates the quest for future contributions that exploit unlabelled data in CTD, compensating the high cost of expert knowledge in this field.

\bibliographystyle{IEEEtran}
\bibliography{IEEEabrv, references, references_ssl}

\appendices

\section{Experimental Testbed}
\label{app:testbed}
All our experiments are performed on a machine equipped with an Intel Xeon W-2195 CPU with 36 cores, 256GB RAM, 2TB SSD NVMe, and Nvidia Titan RTX GPU. The implementation leverages Python3 and the well-known ML library of scikit-learn. The specific ML algorithm used as base for all our models is the RandomForestClassifier. Training such classifier can be parallelized: in particular, we set the \textit{njobs} parameter to use 34 cores (out of 36) of our CPU.

\subsection{Data Preprocessing}
\label{sapp:preprocessing}
\textbf{NID datasets.}
The considered NID datasets all contain more than one malicious class---which we treat differently.
\begin{itemize}
    \item \dataset{CTU13} is composed of 13 PCAP traces, each one containing either benign or malicious traffic belonging to a given botnet family out of 7 possible families. Such traces are transformed into NetFlows via Argus. To prevent overfitting, we remove the full IP-address (we differentiate between internal/external IPs); we also derive some additional metrics (e.g., bytes per second, or the IANA port categories). The entire feature set is provided in Table~\ref{tab:features_ctu}, which is similar to the one in~\cite{apruzzese2020hardening}. Then, we merge all sets according to the specific botnet family, obtaining 7 sets containing either benign samples, or malicious samples belonging to a single family. We exclude those botnet families with less than 1K samples.
    
    \item \dataset{IDS17} is provided as NetFlows, each separated in benign and malicious samples of different attacks. We use the entire feature set provided by the authors of \dataset{IDS17}, but we apply the same internal/external differentiation as done for \dataset{CTU13} to prevent overfitting on individual IP addresses. We aggregate the sets of NetFlows containing DoS attacks into a single set, and we exclude the families underrepresented. We create a dedicated set for the benign samples, which we use for all the experiments.

    \item \dataset{UNB15} the procedure is similar to \dataset{IDS17}. We use the same feature sets provided by the authors (with the usual distinction between internal/external machines). We aggregate the most represented families in 5 distinct sets. 
\end{itemize}

\begin{table}[!h]
    \centering
    \caption{Features considered for \dataset{CTU13}.}
    \resizebox{0.4\columnwidth}{!}{
        \begin{tabular}{ccc}
            \toprule
            \textbf{\#} & \textbf{Feature Name} & \textbf{Type}\\ 
            \midrule
            1 & SrcIP internal  & Bool \\
            2 & DstIP internal & Bool\\
            3 & SrcPort type & Cat\\
            4 & DstPort type & Cat\\
            5 & SrcPort & Cat\\
            6 & DstPort & Cat\\
            7 & Flow Duration $[s]$ & Num \\
            8 & Flow Direction & Bool\\
            9 & Inc-Bytes & Num\\
            10 & Out-Bytes & Num\\
            11 & Total Bytes& Num\\
            12 & Incoming Packets & Num\\
            13 & Outgoing Packets & Num\\
            14 & Total Packets & Num\\
            15 & Bytes per Packet & Num\\
            16 & Bytes per Second & Num\\
            17 & Packets per Second & Num\\
            18 & Inc- per Out-Bytes & Num\\
            \bottomrule
        \end{tabular}
    }
    \label{tab:features_ctu}
\end{table}

The complete break down of the benign and malicious sets for each NID dataset is provided in Table~\ref{tab:breakdown}.

\begin{table}[!h]
    \centering
    \caption{Breakdown of traces for NID datasets.}
    \resizebox{0.7\columnwidth}{!}{
        \begin{tabular}{c?c|c|c|c}
            \toprule
            \textbf{Dataset} & Trace & \begin{tabular}{c} Attack \\ Class \end{tabular} & \begin{tabular}{c} Malicious \\ Samples \end{tabular} $\mathbb{D}_m$ & \begin{tabular}{c} Benign \\ Samples \end{tabular} $\mathbb{D}_b$ \\ 
            \midrule
            
            \multirow{6}{*}{\dataset{CTU13}}
            & 1 & Neris & 246889 & 6473336 \\ \cline{2-5}
            & 2 & e4f8 & 40904 & 2014041 \\ \cline{2-5}
            & 3 & svchost & 4630 & 554284 \\ \cline{2-5}
            & 4 & qvodset & 6127 & 2948062 \\ \cline{2-5}
            & 5 & NSISay & 2168 & 323300 \\ \cline{2-5}
            & 6 & qvodset & 143918 & 7104673 \\ 
            \midrule
            
            \multirow{4}{*}{\dataset{UNB15}}
            & 1 & DoS & 16353 & \multirow{4}{*}{2218756} \\ \cline{2-4}
            & 2 & Exploits & 44525 & \\ \cline{2-4}
            & 3 & Fuzzers & 24246 & \\ \cline{2-4}
            & 4 & Recon & 13987 & \\ 
            \midrule
            
            \multirow{5}{*}{\dataset{IDS17}}
            & 1 & DoS & 252661 & \multirow{5}{*}{2273097} \\ \cline{2-4}
            & 2 & DDoS & 128027 & \\ \cline{2-4}
            & 3 & PortScan & 158930 & \\ \cline{2-4}
            & 4 & SshFtp & 13835 & \\ \cline{2-4}
            & 5 & WebAttack & 2180 & \\ 
            
            \bottomrule
        \end{tabular}
    }
    \label{tab:breakdown}
\end{table}

In all experiments involving NID datasets, we devise ensemble of classifiers, each focused on a specific family (as done in~\cite{apruzzese2020hardening, zhang2021network}). The results presented in our graphs are the average of all classifiers.

\textbf{MD datasets.}
The task of MD has been tackled either as a binary classification~\cite{galen2021empirical} or as a multi-classification ML problem~\cite{Chakraborty:EC2}---the latter focusing on identifying the specific malware family of a given sample. Because \dataset{EMBER} only has a single malicious family, all our MD are binary classifiers. To this purpose, we observe that \dataset{DREBIN} (collected from 2010 to 2012) has hundreds of malware families, some of which extremely underrepresented: hence, we considered the samples of top10 families and treated them as a single class. On the other hand, \dataset{AndMal20} is more balanced across families, hence we consider all samples as a single malicious class.

\textbf{PWD datasets.}
All PWD datasets include malicious samples of a single class. To the best of our knowledge, PWD has always been treated as a binary classification problem; hence, developing our PWD is more straightforward compared to NID and MD.
The \dataset{UCI} and \dataset{Mendeley} datasets are provided directly as features; whereas for \dataset{$\delta$Phish}, we extract the most significant features based on the information of each provided webpage (HTML, URL, DNS) to achieve a similar feature set as the other two~\cite{mohammad2014predicting}.
The queries to the DNS servers were performed in 2019, and some returned null results; we still consider these pages in our evaluation, as this happened for both benign and malicious pages. We provide in Table~\ref{tab:features_deltaphish} the complete feature set used for \dataset{$\delta$Phish}.

\begin{table}[htpb!]
\centering
\caption{Features computed for the \dataset{$\delta$Phish} dataset.}
\resizebox{0.6\columnwidth}{!}{%
        \begin{tabular}{|c|c|c|}
            \hline
            \textit{URL-features} & \textit{REP-features} & \textit{HTML-features} \\
            \hline
            \hline
            IP address & SSL final state & SFH  \\ \hline
            '@' (at) symbol & URL/DNS mismatch & Anchors \\ \hline
            '-' (dash) symbol & DNS Record & Favicon \\ \hline
            Dots number & Domain Age & iFrame  \\ \hline
            Fake HTTPS & PageRank & MailForm  \\ \hline
            URL Length & PortStatus & Pop-Up  \\ \hline
            Redirect & Redirections & RightClick \\ \hline
            Shortener &  & Objects \\ \hline
            dataURI &  & StatusBar \\ \hline
            &  & Meta-Scripts \\ \hline
            &  & CSS \\ \hline
        \end{tabular}
}
\label{tab:features_deltaphish}
\end{table}

\subsection{Developing the SsL models}
\label{app:ssl-methods}
We describe our SsL methods and provide an example.

\textbf{`Pure' pseudo-labelling}. We first use \smallmath{SL} to predict the pseudo-labels of $\mathbb{U}$ and provide the confidence, $c$, of such predictions. Depending on $c$ (see §\ref{ssec:approaches}), the pseudo-labelled samples will be inserted in $\mathbb{L}$, resulting in a `mixed' $\mathbb{L}$ used to train a `pure' pseudo-labelling model. For the pseudo-labelling with retraining, we use \smallmath{\pi\,SsL} to predict the labels of the remaining samples in $\mathbb{U}$, and assign those with $c\geq99\%$ to $\mathbb{L}$ (with the pseudo-label).

\textbf{`Pure' active learning}.
CEF-SsL composes another $\mathbb{L}$ (and another $\mathbb{U}$) with a two-step approach, by assuming that \textit{half} of the labelling budget is used initially, and the remaining half is used for labelling the suggested samples. To this purpose, the initial $\mathbb{L}$ is changed by randomly removing half of its benign and malicious samples, therefore restoring the budget $\mathcal{L}$ to half of its initial value. Such $\mathbb{L}$ is used to train a `support' SL model that predicts the labels of $\mathbb{U}$ and the corresponding confidence, $c$. CEF-SsL then simulates a human oracle that assigns the correct label to the samples that meet a confidence threshold (cf. §\ref{ssec:approaches}) by accounting for such costs from the residual $\mathcal{L}$. Such samples, with their correct label, will be inserted in $\mathbb{L}$. 
To allow a fair comparison where all the models use an $\mathbb{L}$ with the same size, we assume that the cost for labelling each `suggested' sample is standardized. This assumption is realistic, because it assumes that the sample already exists (it comes from $\mathbb{U}$) and a human operator can more efficiently verify a sample that is being provided with, compared to choosing and verifying samples randomly, or entirely creating new ones. The resulting new $\mathbb{L}$ (containing only correct labels) is then used to train the corresponding active learning model. 

\textbf{Pseudo-active models.}
We use the `support' SL (trained on the `halved' $\mathbb{L}$) to predict the pseudo labels of $\mathbb{U}$, and put all those with $c\,\geq\,99\%$ in $\mathbb{L}$ (with the pseudo label). Such `mixed' $\mathbb{L}$ is used to train a support \smallmath{\pi\,SsL}, which predicts the remaining $\mathbb{U}$: the samples whose confidence meets the criteria in §\ref{ssec:approaches} are randomly put into the `mixed' $\mathbb{L}$ until finishing the leftover budget. Such $\mathbb{L}$ (having the initial correct labels, the pseudo labels, and the `suggested' correct labels) is used for training the pseudo-active SsL model.

\textbf{Example.}
Consider the unbalanced case on \dataset{CTU13} where $\mathcal{C}_m$=$2\mathcal{C}_b$, and $\mathcal{L}$=200 (cf. Table~\ref{tab:scenarios}). For the `pure' pseudo models and the baseline \smallmath{SL}, the $\mathcal{L}$ is used all at the beginning, and the final $\mathbb{L}$ will always have 150 correctly labelled samples: 50 malicious samples and 100 benign.
For active learning methods, the first half of the labelling budget (100) is used for the initial learner, by randomly choosing and removing 50 benign samples and 25 malicious samples from the previously randomly drawn $\mathbb{L}$; doing so leads to a smaller initial $\mathbb{L}$ with 75 samples. Then, because of our standardized assumption, the oracle will randomly assign the correct label to 75 samples that meet the desired confidence criteria, irrespective of their class. The oracle will \textit{not} label more samples than what it is allowed by the budget. Nonetheless, the corresponding model will be tested on $\mathbb{F}$, and then the entire process is repeated 5 times to account for randomness in choosing the `suggested' samples.

\section{Benchmark}
\label{app:benchmark}

We present our benchmark evaluation, whose nature is \textit{exploratory}: we are not interested in providing results that `outperform' the state-of-the-art. Our focus is providing the first statistically validated benchmark for SsL methods in CTD and promote future analyses. Hence, we consider realistic scenarios where the amount of labelled data is scarce: in our evaluation, we never use more than 2.4K labelled samples. As a consequence, some results may appear to be underwhelming: we consider such outcomes to be \textit{positive}, as they highlight the huge improvement margin concealed by SsL.

\begin{figure*}[!htbp]
     \centering
     \begin{subfigure}[b]{0.9\textwidth}
         \centering
         \includegraphics[width=\columnwidth]{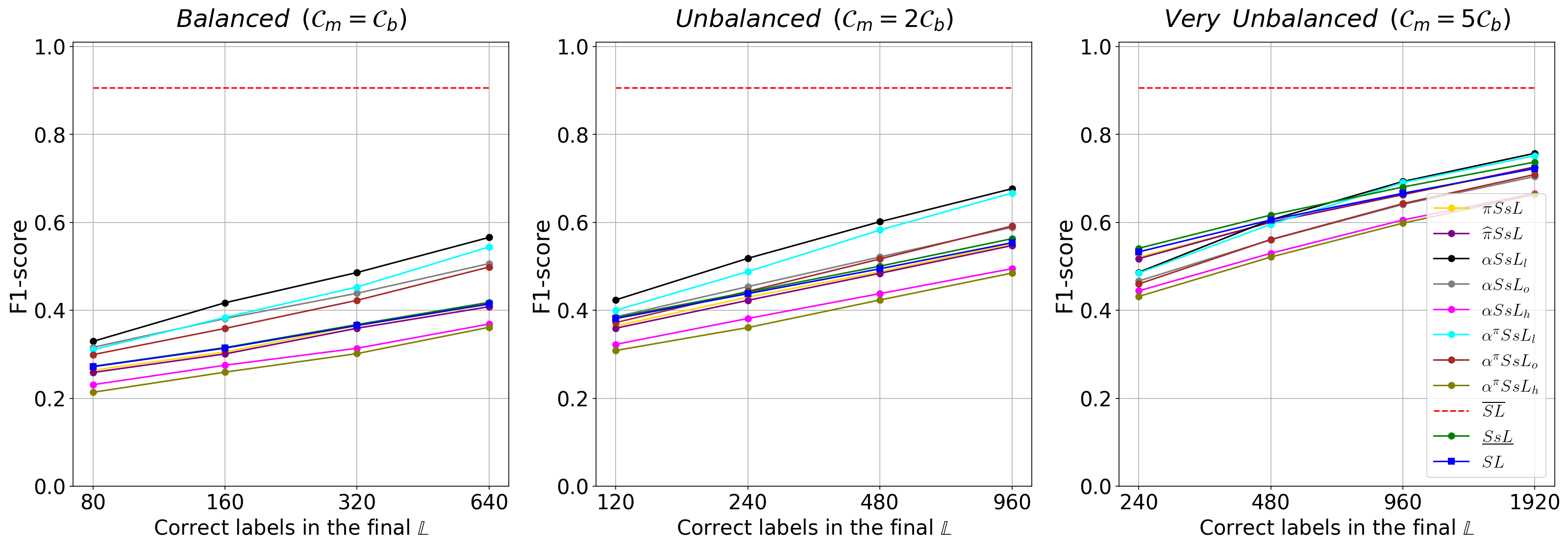}
         \caption{Results on \dataset{\textbf{DREBIN}}. For every plot, each `point' represents the average results of 100 models (and 5 times as much for all models using active learning).}
         \label{sfig:drebin}
     \end{subfigure}
     \begin{subfigure}[b]{0.9\textwidth}
         \centering
         \includegraphics[width=\columnwidth]{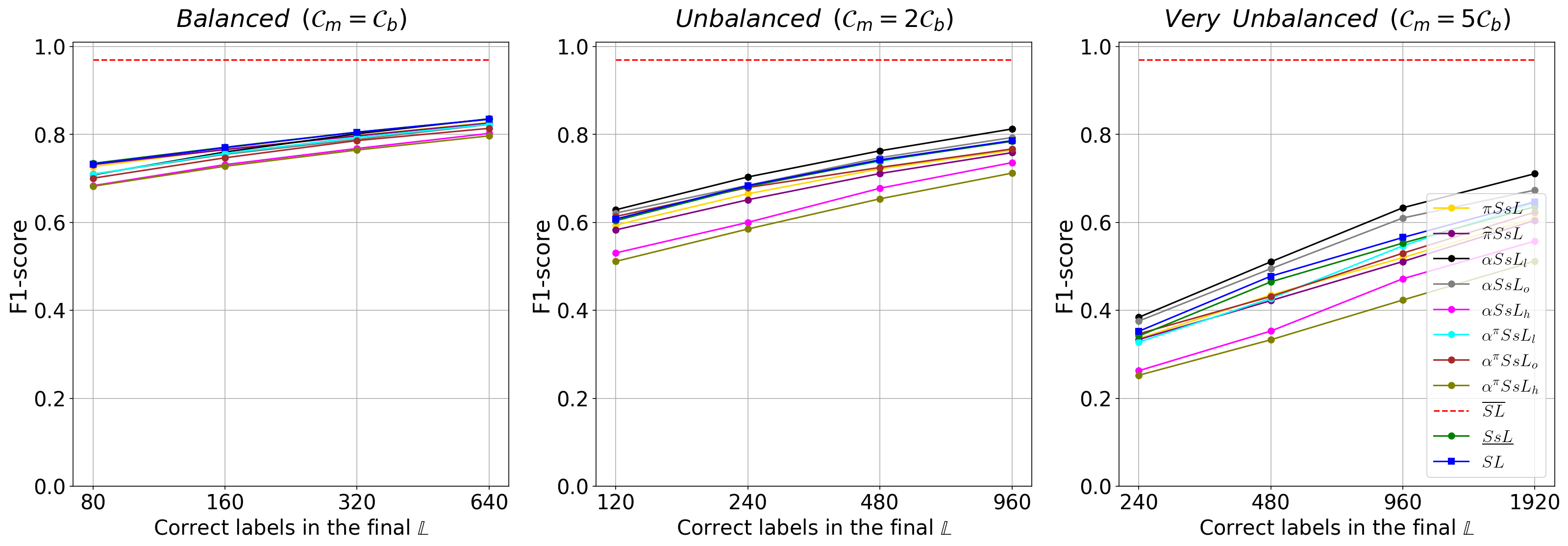}
         \caption{Results on \dataset{\textbf{Ember}}. For every plot, each `point' represents the average results of 100 models (and 5 times as much for all models using active learning).}
         \label{sfig:ember}
     \end{subfigure}
     \begin{subfigure}[b]{0.9\textwidth}
         \centering
         \includegraphics[width=\columnwidth]{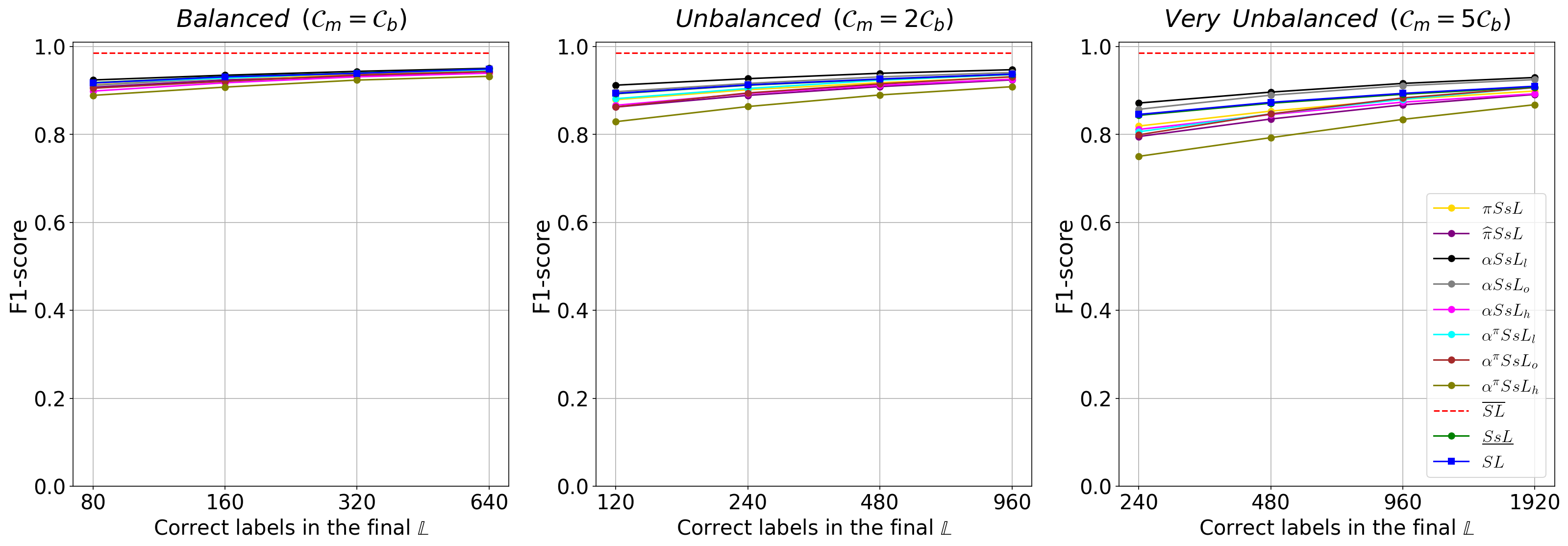}
         \caption{Results on \dataset{\textbf{AndMal20}}. For every plot, each `point' represents the average results of 100 models (and 5 times as much for all models using active learning).}
         \label{sfig:andmal}
     \end{subfigure}
     \caption{\textbf{Malware Detection}. For each dataset, we report the results for the three cost (or balancing) scenarios. Each scenario is shown in a plot, where the y-axis reports the F1-score and the x-axis the (increasing) labelling budget. Each method is denoted with a line on each plot.}
    \label{fig:results_MD}
\end{figure*}

\textbf{Results.} The overall benchmark results are shown in Figs.~\ref{fig:results_nid} for NID datasets; Figs~\ref{fig:results_pd} for PWD datasets; and Figs~\ref{fig:results_MD} for MD datasets. Each figure consists in a set of 3 subfigures, each focused on a specific dataset. Every subfigure reports 3 plots, each focused on a specific `balance' scenario (i.e., a specific $\mathcal{C}$). Every plot reports the F1-score (vertical axis) achieved by all the considered SsL methods (lines) for increasing labelling budgets $\mathcal{L}$ (horizontal axis). We observe that the performance of our MD for \dataset{DREBIN} (Fig.~\ref{sfig:drebin}) is significantly worse than on \dataset{Ember} (Fig.~\ref{sfig:ember}) and on \dataset{AndMal20} (Fig.~\ref{sfig:andmal}). Such phenomenon is due to our chosen aggregation strategy. Indeed, the considered malicious samples in \dataset{DREBIN} belong to 10 different families; however, such samples are randomly chosen when composing the $\mathbb{L}$, i.e., $\mathbb{L}$ (which is very small) may not contain some families. As such, if these families are notably different from those included in $\mathbb{F}$ and---at the same time---vastly present in $\mathbb{F}$, then the MD will exhibit low performance. This phenomenon, however, does not appear in \dataset{AndMal20} (for which we also aggregate families into a single class): an explanation is that the malicious families in \dataset{AndMal20} are more similar to each other than those in \dataset{DREBIN}.

\textbf{How many experiments does the benchmark include?} We report in Table~\ref{tab:classifiers} the overall number of models considered in our evaluation. Specifically, for each dataset we report $n,k$ and the corresponding N. Multiplying all of these numbers yields the `runs' of any model that does not leverage active learning, i.e., \smallmath{\overbar{SL}}, \smallmath{SL}, \smallmath{\underline{SsL}}, \smallmath{\pi\,SsL}, \smallmath{\widehat{\pi}\,SsL}. Because we draw the samples for active learning randomly and we repeat such draw 5 times for each corresponding model, all the 6 active learning models (i.e., \smallmath{\alpha\,SsL_h}, \smallmath{\alpha\,SsL_o}, \smallmath{\alpha\,SsL_l}, \smallmath{\alpha^{\pi}\,SsL_h}, \smallmath{\alpha^{\pi}\,SsL_o}, \smallmath{\alpha^{\pi}\,SsL_l}) are assessed 5 times as much. Hence, for every line in a given plot, each `point' is the average results of as many models as reported in Table~\ref{tab:classifiers}. As an example, on \dataset{CTU13}, each point in a given plot of Fig.~\ref{sfig:ctu13} is the average F1-score of 990 models if the line is related to an active learning method, or 198 models if not---all of which evaluated for the corresponding value of $\mathcal{L}$ and cost scenario $\mathcal{C}$.

\begin{table}[!htbp]
    \centering
    \caption{Total amount of results included in each `point' of Figs~\ref{fig:results_nid}--\ref{fig:results_MD}.}
    \label{tab:classifiers}
    \resizebox{0.8\columnwidth}{!}{
        \begin{tabular}{c?c|c|c|c|c}
             \begin{tabular}{c}
                  \textbf{CTD} \\
                  (Figure)
             \end{tabular}
             & \begin{tabular}{c}
                  \textbf{Dataset} \\
                  (Subfigure)
             \end{tabular}
             & $n,k$ & N
             & {\scriptsize\begin{tabular}{c}
                  Active \\
                  models (6) 
             \end{tabular}}
             & {\scriptsize\begin{tabular}{c}
                  Other \\
                  models (5) 
             \end{tabular}}
             \\ 
             
             \toprule
             
             \multirow{3}{*}{{
             \begin{tabular}{c}
                  \textbf{NID}  \\
                  Figs.~\ref{fig:results_nid} 
             \end{tabular}
             }}
              & 
            \dataset{CTU13} (Fig.~\ref{sfig:ctu13}) & (11,3) & 6 & 990 & 198 \\
            & \dataset{UNB15} (Fig.~\ref{sfig:unb15}) & (23,4) & 4 & 1840 & 368 \\
            & \dataset{IDS17} (Fig.~\ref{sfig:ids17}) & (15,3) & 5 & 1125 & 225 \\

              \midrule
             
              \multirow{3}{*}{{
             \begin{tabular}{c}
                  \textbf{PWD}  \\
                  Figs.~\ref{fig:results_pd} 
             \end{tabular}
             }}
              & 
            \dataset{UCI} (Fig.~\ref{sfig:uci}) & (20,5) & 1 & 500 & 100 \\
            & \dataset{Mendeley} (Fig.~\ref{sfig:mendeley}) & (20,5) & 1 & 500 & 100\\
            & \dataset{$\delta$Phish} (Fig.~\ref{sfig:deltaphish}) & (20,5) & 1 & 500 & 100 \\
              
              \midrule
              
              \multirow{3}{*}{{
             \begin{tabular}{c}
                  \textbf{MD}  \\
                  Figs.~\ref{fig:results_MD} 
             \end{tabular}
             }}
              & 
            \dataset{Drebin} (Fig.~\ref{sfig:drebin}) & (20,5) & 1 & 500 & 100 \\
            & \dataset{Ember} (Fig.~\ref{sfig:ember}) & (20,5) & 1 & 500 & 100 \\
            & \dataset{AndMal20} (Fig.~\ref{sfig:andmal}) & (20,5) & 1 & 500 & 100 \\
            
            \bottomrule
        \end{tabular}

    }
\end{table}

All the values in Table~\ref{tab:classifiers} correspond \textit{only} to a specific combination of $\mathcal{L}$ and $\mathcal{C}$, meaning that the overall amount of models developed in our evaluation is 12 times as much---for each \smallmath{\mathbb{D}}. In summary, the results of SoK paper correspond to $500\,760$ active learning models and $83\,460$ non-active learning models, for a total of $584\,220$ models.

\section{Case Study: Comparison with a prior work}
\label{app:casestudy}
Let us discuss a case study where we compare our evaluation with a recent work sharing a similar testbed.

The combination of pseudo labelling and active learning has been investigated also by Zhang et al.~\cite{zhang2021network} on \dataset{CTU13} and \dataset{IDS17} by using--among others--the same confidence levels as our implementation: above $99\%$ for the pseudo labelling, and below $1\%$ for active learning (making it equivalent to our \smallmath{\pi\,SsL} and \smallmath{\alpha\,SsL_l}).
In~\cite{zhang2021network}, the two source datasets (\dataset{IDS17} and \dataset{CTU13}) are divided into a single set of benign samples; whereas the malicious samples are distributed into several sets depending on their attack family, and then aggregated into a single malicious class. This is a valid operation, but it slightly differs from our testbed, because we treat malicious classes separately for each NID dataset.

Let us describe the evaluation methodology of~\cite{zhang2021network}. Our first observation is that, after `preprocessing' the source datasets, the authors of~\cite{zhang2021network} obtain 100K samples for \dataset{CTU13} and 600K samples for \dataset{IDS17}. In contrast, after preprocessing, we obtain 20M samples for \dataset{CTU13} and 3M for \dataset{IDS17}---and this is despite removing some underrepresented families (cf. Table~\ref{tab:datasets}). We are not aware of the reason of this gap, but also other studies (e.g.,~\cite{panigrahi2018detailed, apruzzese2020hardening}) obtain similar compositions as ours.

Regardless, each preprocessed dataset in~\cite{zhang2021network} is further filtered to obtain a $\mathbb{D}$ containing 15K benign samples and 9K malicious samples. Such $\mathbb{D}$ (which represents only a small portion of their preprocessed traffic, i.e. 10\% for \dataset{CTU13}, and 4\% for \dataset{IDS17}) is \textit{never} changed in the evaluation and the remaining samples are \textit{never} used. In contrast, we consider as $\mathbb{D}$ the entire datasets after preprocessing.

Then, the authors of~\cite{zhang2021network} partition such $\mathbb{D}$ into $\mathbb{\overbar{L}}$ and $\mathbb{F}$ by using a 70:30 split (we use 80:20), resulting in a $\mathbb{\overbar{L}}$ with 17K samples, and a $\mathbb{F}$ with 7K samples---both having $\rho$=(62,38).
The split is done randomly, but the process is never repeated and the $\mathbb{F}$ stays the same for the entire evaluation. What would have happened if $\mathbb{F}$ contained different samples? We address this issue by changing $\mathbb{F}$ multiple times ($k$) for each malicious class (N), meaning 18 times for \dataset{CTU13}, and 15 times for \dataset{IDS17}. 

To obtain their $\mathbb{L}$ (and corresponding $\mathbb{U}$), the authors of~\cite{zhang2021network} isolate a variable portion of samples from $\mathbb{\overbar{L}}$, specifically either 5\%, 10\% or 20\% (hence, the correct labels in their $\mathbb{L}$ range from $\sim$1K to $\sim$4K). The choice of samples put in $\mathbb{L}$ is done randomly in~\cite{zhang2021network}, but the experiments are repeated only 10 times. In contrast, we do so $n$ times for each new $\mathbb{F}$, meaning that we do so 198 times for \dataset{CTU13} and 225 times for \dataset{IDS17}. This increases the confidence of our results. Moreover, we also consider different balance ratios, whereas the balancing in~\cite{zhang2021network} is always a fixed $\rho$=(62,38) for all sets.

Finally, in~\cite{zhang2021network} they consider the \smallmath{SL} baseline, but neglect the \smallmath{\underline{SsL}} baseline and the \smallmath{\overbar{SL}} baseline. Both lacks are significant: without \smallmath{\underline{SsL}}, it is not possible to estimate the least possible benefit provided by $\mathbb{U}$ (if any); without \smallmath{\overbar{SL}}, it is not possible to determine any upper bound in performance. If the \smallmath{SL} is not far from \smallmath{\overbar{SL}}, then it may not be worth in investing in $\mathcal{U}$ for using SsL methods. 

\begin{figure*}[!htbp]
     \centering
     \begin{subfigure}[b]{0.9\textwidth}
         \centering
         \includegraphics[width=\columnwidth]{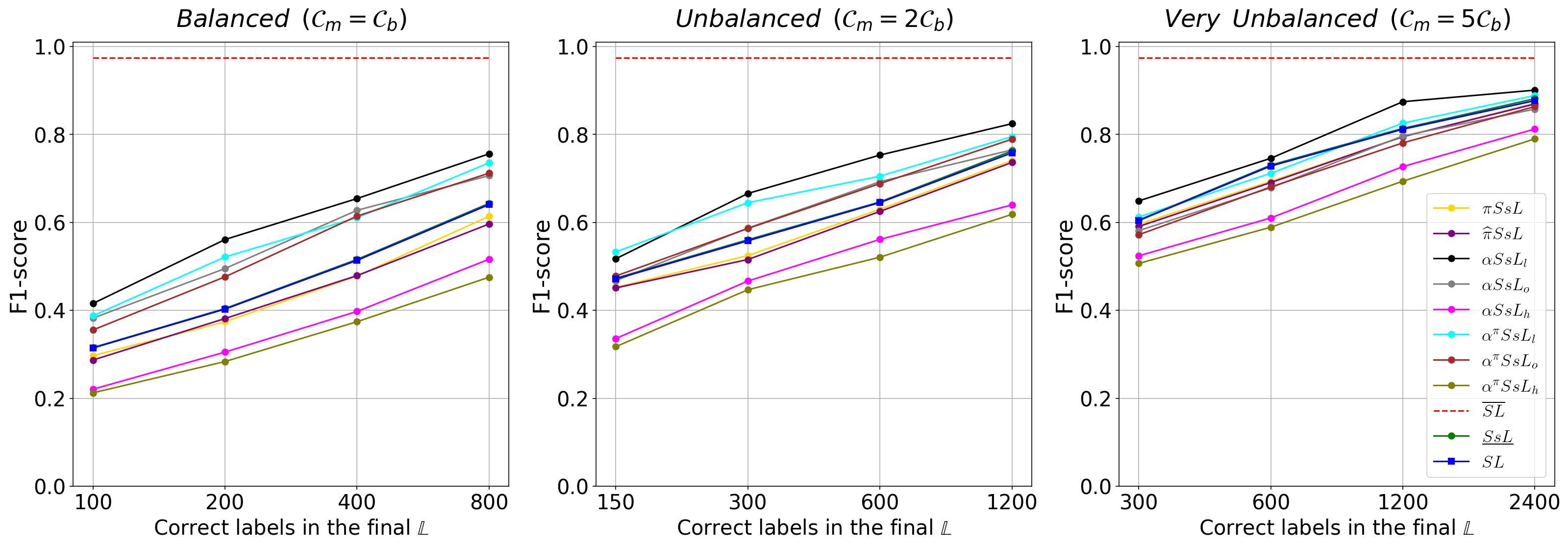}
         \caption{Results on \dataset{\textbf{CTU13}}. For every plot, each `point' represents the average results of 198 models (and 5 times as many for all models using active learning).}
         \label{sfig:ctu13}
     \end{subfigure}
     \begin{subfigure}[b]{0.9\textwidth}
         \centering
         \includegraphics[width=\columnwidth]{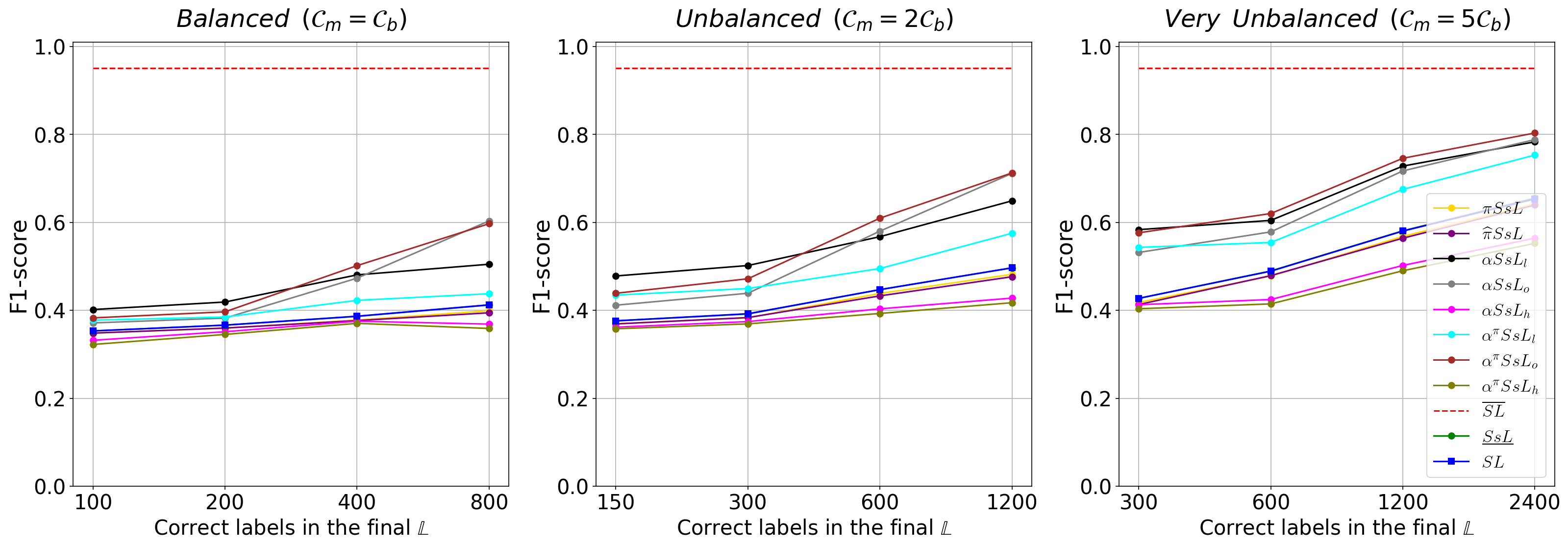}
         \caption{Results on \dataset{\textbf{UNB15}}. For every plot, each `point' represents the average results of 368 models (and 5 times as much for all models using active learning).}
         \label{sfig:unb15}
     \end{subfigure}
     \begin{subfigure}[b]{0.9\textwidth}
         \centering
         \includegraphics[width=\columnwidth]{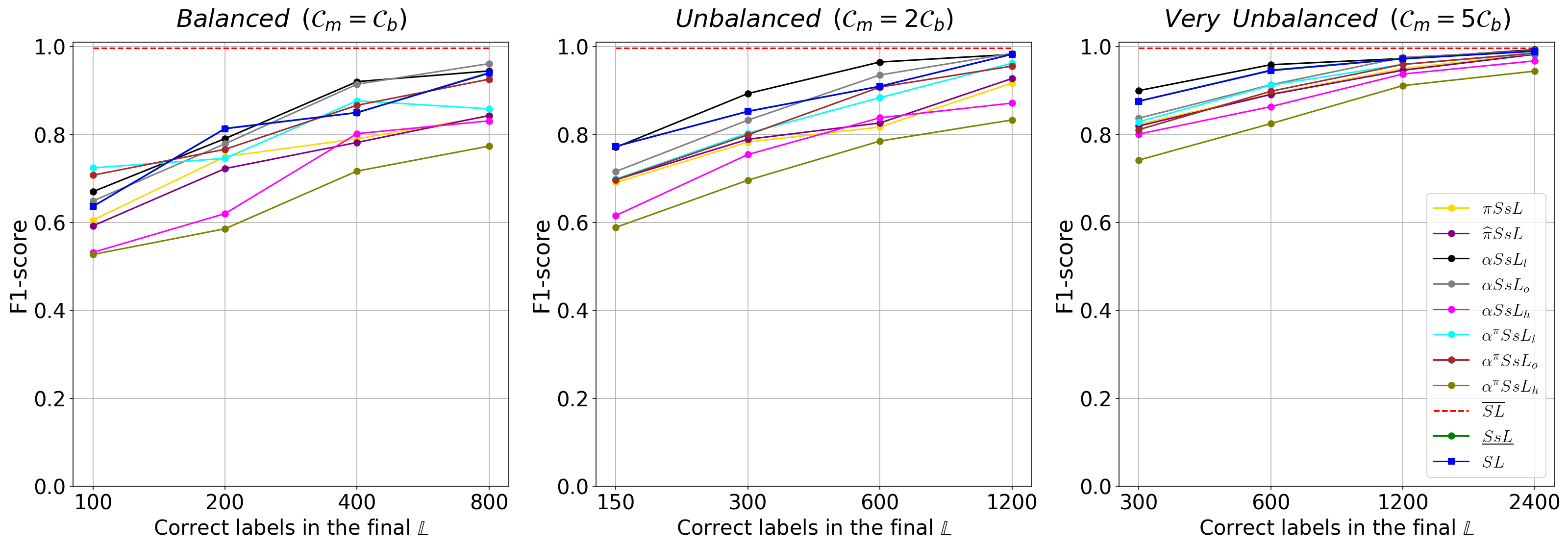}
         \caption{Results on \dataset{\textbf{IDS17}}. For every plot, each `point' represents the average results of 225 models (and 5 times as much for all models using active learning).}
         \label{sfig:ids17}
     \end{subfigure}
     \caption{\textbf{Network Intrusion Detection}. For each dataset, we report the results for the three cost (or balancing) scenarios. Each scenario is shown in a plot, where the y-axis reports the F1-score and the x-axis the (increasing) labelling budget. Each method is denoted with a line on each plot.}
    \label{fig:results_nid}
\end{figure*}

Let us compare our results, with the aim of pinpointing what `issues' prevent deriving actionable conclusions on the impact of unlabelled data in~\cite{zhang2021network}. For simplicity, we focus on the F1-score achieved on \dataset{CTU13} (cf. Fig.\ref{sfig:ctu13}).
The baseline \smallmath{SL} in~\cite{zhang2021network} achieves a lower performance than ours despite being trained on more samples: the one in~\cite{zhang2021network} obtains 0.81 F1-score when trained on 4000 samples, whereas our baseline \smallmath{SL} trained with an $\mathbb{L}$ of 2400 samples (the highest we considered) reached 0.87 F1-score, as evidenced by the rightmost plot in Fig.\ref{sfig:ctu13}. 
When applying pseudo-labelling in~\cite{zhang2021network}, the performance increases from 0.81 to 0.83 F1-score. All these results contrast with ours, because the performance of our corresponding model, \smallmath{\pi\,SsL}, is lower than the baseline. However, while our results take into account a total of 198 trials, the ones in~\cite{zhang2021network} are performed only 10 times, which is hardly enough to make any informed decision on whether it is truly convenient to invest in $\mathcal{U}$.

Finally, the authors of~\cite{zhang2021network} do not maintain the original $\mathcal{L}$ when applying active learning, and do not report how many samples require to be `actively labelled' as they inject all those within the confidence threshold (below 1\%) into $\mathbb{L}$ (as also done in Tesseract~\cite{pendlebury2019tesseract}), making any comparison with our models (and, also, with their baselines) unfair.

We can conclude that the evaluation performed in~\cite{zhang2021network} can only show that, under certain conditions, some SsL methods can improve the performance. However, the results obtained cannot certify that such improvement is significant, because they are conducted in a fixed setting (same $\mathbb{F}$, same balance ratio, only 10 runs), despite being performed on two datasets. Hence, the question ``do I need $\mathbb{U}$?'' is still open.

\section{CEF-SsL for multi-classification}
\label{app:multiclass}
In our evaluation we treat CTD as a binary classification problem, where a sample is either benign or malicious. The motivation is that N+1 classification (with N$>$1) assumes a `closed world' scenario where all malicious classes are known beforehand, which is hardly the case in the dynamic cybersecurity landscape. Nevertheless, some specific applications may favor SsL method devoted to multi-classification: applying CEF-SsL in similar settings is possible, i.e., by manually specifying the cost to label \textit{each} malicious sample $\mathcal{C}_x$, and performing hundreds of runs of CEF-SsL---but this is empirically difficult from a research perspective. Some alternatives exist, but they are also challenging.

The main challenge is \textit{randomly choosing a limited amount of samples from N malicious classes}. For instance, some datasets (e.g., \dataset{DREBIN}) only have a very limited number of samples for some classes; a possible workaround is applying some aggregation techniques and create some `macro' classes; however, such approach may introduce some experimental bias.
Another option is removing those classes for which only few samples are available: in this case, however, the ML model may perform poorly if such families `appear' after the model is deployed. To mitigate such problem, all underrepresented classes can be merged into a dedicated `other' class: the ML model may retain some performance at inference; but it may also be confused when the more represented classes present similarities with the samples of the `other' class.

Another challenge is \textit{composing `appropriate'} partitions, i.e., $\mathbb{L}$, $\mathbb{U}$ and $\mathbb{F}$. It is well-known that, in reality, benign events are more abundant than malicious ones, and it is common to compose train/test partitions where benign samples are the majority. However, when malicious samples belong to different classes and the labeling budget is limited, it begs the question of ``how many samples \textit{per class} should be included in $\mathbb{L}$?''. As an example, assume that $\mathcal{L}$=500; how should $\mathbb{L}$ be composed when a dataset contains 10K benign samples, alongside three malicious classes, the first with 2000 samples, the second with 100 samples, and the third with 50 samples? And what about $\mathbb{U}$ and $\mathbb{F}$?
A possibility is using the relative distribution in a given dataset, but it may induce bias or skew the model into favoring the majority class. Conversely, it is possible to infer which family is more popular `in the wild', but this may require extra resources to study updated security feeds.

\end{document}